\documentclass[twoside,twocolumn,9pt]{article}
\usepackage{extsizes}
\usepackage[super,sort&compress,comma]{natbib} 
\usepackage[version=3]{mhchem}
\usepackage[left=1.5cm, right=1.5cm, top=1.785cm, bottom=2.0cm]{geometry}
\usepackage{balance}
\usepackage{mathptmx}
\usepackage{sectsty}
\usepackage{graphicx} 
\usepackage{lastpage}
\usepackage[format=plain,justification=justified,singlelinecheck=false,font={stretch=1.125,small,sf},labelfont=bf,labelsep=space]{caption}
\usepackage{float}
\usepackage{fancyhdr}
\usepackage{fnpos}
\usepackage[english]{babel}
\addto{\captionsenglish}{%
  
}
\usepackage{array}
\usepackage{droidsans}
\usepackage{charter}
\usepackage[T1]{fontenc}
\usepackage[usenames,dvipsnames]{xcolor}
\usepackage{setspace}
\usepackage[compact]{titlesec}

\usepackage{pslatex}

\usepackage{booktabs} 
\usepackage{ulem}
\usepackage{xr}
\usepackage{indentfirst}
\usepackage{gensymb} 
\usepackage{textcomp} 
\usepackage{tabularx,multirow}

\makeatletter
\newcommand*{\addFileDependency}[1]{
  \typeout{(#1)}
  \@addtofilelist{#1}
  \IfFileExists{#1}{}{\typeout{No file #1.}}
}
\makeatother

\makeFNbottom
\makeatletter
\renewcommand\LARGE{\@setfontsize\LARGE{15pt}{17}}
\renewcommand\Large{\@setfontsize\Large{12pt}{14}}
\renewcommand\large{\@setfontsize\large{10pt}{12}}
\renewcommand\footnotesize{\@setfontsize\footnotesize{7pt}{10}}
\makeatother

\makeatletter 
\renewcommand\@biblabel[1]{#1}            
\renewcommand\@makefntext[1]%
{\noindent\makebox[0pt][r]{\@thefnmark\,}#1}
\makeatother 

\sectionfont{\sffamily\Large}
\subsectionfont{\normalsize}
\subsubsectionfont{\bf}
\titlespacing*{\section}{0pt}{4pt}{4pt}
\titlespacing*{\subsection}{0pt}{15pt}{1pt}

\definecolor{cream}{RGB}{222,217,201}

\makeatletter 
\newlength{\figrulesep} 
\makeatother
\begin{document}

\twocolumn[
  \begin{@twocolumnfalse}
\vspace{1em}
\sffamily
\begin{tabular}{m{0cm} p{18cm} }
& \noindent\LARGE{\textbf{Combinatorial synthesis of cation-disordered manganese tin nitride \ce{MnSnN2} thin films with magnetic and semiconducting properties}} \\
\vspace{0.3cm} & \vspace{0.3cm} \\

 & \noindent\large{Christopher L. Rom,\textit{$^{a, b}$} 
 Rebecca W. Smaha,\textit{$^{a}$} 
 Celeste L. Melamed,\textit{$^{a, c}$}
 Rekha R. Schnepf,\textit{$^{a, c}$}
 Karen N. Heinselman,\textit{$^{a}$} 
 John S. Mangum,\textit{$^{a}$} 
 Sang-Jun Lee,\textit{$^{d}$}
 Stephan Lany,\textit{$^{a}$}
 Laura T. Schelhas,\textit{$^{a, d}$}
 Ann L. Greenaway,\textit{$^{a}$}
 James R. Neilson,\textit{$^{b, e}$}
 Sage R. Bauers,\textit{$^{a}$}
 Jennifer S. Andrew,$^{\ast}$\textit{$^{a,f}$}
 Adele C. Tamboli\textit{$^{a, c}$}} \\

 & \noindent\normalsize{Magnetic semiconductors may soon improve the energy efficiency of computers, but materials exhibiting these dual properties remain underexplored. 
Here, we report the computational prediction and realization of a new magnetic and semiconducting material, \ce{MnSnN2}, via combinatorial sputtering of thin films.  
Grazing incidence wide angle X-ray scattering and laboratory X-ray diffraction studies show a wide composition tolerance for this wurtzite-like \ce{MnSnN2}, ranging from  $20\% <$ Mn/(Mn+Sn) $< 65$\% with cation disorder across this composition space.
Magnetic susceptibility measurements reveal a low-temperature transition ($T^{\mathrm{*}} \approx 10$~K) for \ce{MnSnN2} and strong antiferromagnetic correlations, although the ordering below this transition may be complex. This finding contrasts with bulk \ce{MnSiN2} and \ce{MnGeN2}, which exhibited antiferromagnetic ordering above 400~K in previous studies.
Spectroscopic ellipsometry identifies an optical absorption onset of 1~eV for the experimentally-synthesized phase exhibiting cation disorder, consistent with the computationally-predicted 1.2~eV bandgap for the cation-ordered structure. 
Electronic conductivity measurements confirm the semiconducting nature of this new phase by showing increasing conductivity with increasing temperature. 
This work adds to the set of known semiconductors that are paramagnetic at room temperature and will help guide future work targeted at controlling the structure and properties of semiconducting materials that exhibit magnetic behavior. 
} \\

\end{tabular}

 \end{@twocolumnfalse} \vspace{0.6cm}

  ]

\renewcommand*\rmdefault{bch}\normalfont\upshape
\rmfamily
\section*{}
\vspace{-1cm}


\footnotetext{\textit{$^{a}$~National Renewable Energy Laboratory, Golden, Colorado 80401, United States}}
\footnotetext{\textit{$^{b}$~Department of Chemistry, Colorado State University, Fort Collins, CO, 80523-1872, United States}}
\footnotetext{\textit{$^{c}$~Physics Department, Colorado School of Mines, Golden, Colorado 80401, United States}}
\footnotetext{\textit{$^{d}$~Stanford Synchrotron Radiation Lightsource, SLAC National Accelerator Laboratory, Menlo Park, California 94025, United States}}
\footnotetext{\textit{$^{e}$~School of Advanced Materials Discovery, Colorado State University, Fort Collins, CO, 80523-1872, United States}}
\footnotetext{\textit{$^{f}$~Department of Materials Science and Engineering, University of Florida, Gainesville, Florida 32611, United States}}
\footnotetext{$^{\ast}$~Email: jandrew@mse.ufl.edu}

\footnotetext{\dag~Electronic Supplementary Information (ESI) available}

\section{Introduction} 

Ternary nitrides are an emerging class of semiconducting materials.\cite{greenaway2021ternaryReview} 
In particular, II-IV-\ce{N2} compounds offer a highly tunable platform that are structurally compatible with well-developed III-N technologies (e.g., GaN).\cite{martinez2017synthesis_II_IV_V2, zakutayev2022experimental} 
In the past few years, many new compounds were discovered in this phase space as thin films (e.g., \ce{MgTiN2}, \ce{ZnMoN2}, \ce{ZnZrN2}),\cite{sun2019map, arca2018redoxZnMoN2, woods2022role} bulk solids (e.g., \ce{CaSnN2}, \ce{CaTiN2}),\cite{kawamura2021synthesisBulkCaSnN2, li2017high_bulkCaTiN2} or both (e.g., \ce{MgSnN2}, \ce{MgZrN2}).\cite{greenaway2020combinatorialMgSnN2, kawamura2020synthesisMgSnN2, bauers2019ternaryRocksalts, rom2021bulkMgZrN2}
These particular compounds---with \ce{Mg^{2+}}, \ce{Ca^{2+}}, and \ce{Zn^{2+}} as the divalent cation---have been investigated for their semiconducting properties. 
However, using transition metals with unpaired electrons---like \ce{Mn^{2+}}---as the divalent cation may add technologically useful magnetic properties. 

For example, magnetic semiconductors are of interest for applications in spintronic devices.\cite{wolf2001spintronics}
By using both charge and spin, spintronics may enable more energy efficient computing than traditional electronics, which rely on charge alone. 
While these devices have largely relied on ferromagnetic materials to record information in spins, ferromagnetic devices face the drawback of memory erasure if accidentally exposed to strong magnetic fields.\cite{gomonay2014AFMspintronics} 
A breakthrough came in 2016, when Wadley, et al. reported the use of an antiferromagnetic material (CuMnAs) to read and write information based on the direction of the anti-aligned spins.\cite{wadley2016electricalSwitchingofAMF_CuMnAs} 
Basing spintronic devices on antiferromagnetic materials comes with the benefit of rapid switching and robust storage, and is a promising direction for increasing the efficiency of computer processing.\cite{baltz2018antiferromagneticSpintronicsReview} 

II-IV-\ce{N2} semiconductors with \ce{Mn^{2+}} offer an underexplored platform for these antiferromagnetic spintronics. 
Bulk methods have been used to synthesize \ce{MnSiN2} and \ce{MnGeN2}, and these phases are known to be antiferromagnetic (AFM) semiconductors with N\'eel temperatures of $T_N = 490$~K and 448~K, respectively.\cite{esmaeilzadeh2006crystalMnSiN2, wintenberger1972etudeMnGeN2} These materials take on a wurtzite-derived structure, with cations and anions in tetrahedral coordination environments. They are structurally analogous to GaN, suggesting the possibility of AFM spintronic integration into III-N based device stacks. In this vein, Liu, et al. reported epitaxial growth of \ce{MnGeN2} films by molecular-beam epitaxy and identified magnetic properties that varied with composition and substrate.\cite{liu2012structural_MBE_MnGeN2} However, much remains to be learned about the fundamental chemistry and physics of these materials.

We note that the tin-containing analogue, \ce{MnSnN2}, has not yet been reported. Thus, the discovery of \ce{MnSnN2} presents an opportunity to deepen our understanding of ternary II-IV-\ce{N2} nitrides in general as well as the physics underlying this \ce{Mn$M$N2} ($M$ = Si, Ge, Sn) wurtzite family in particular. 
With the powerful tool of combinatorial sputtering, we can answer questions like: How much off-stoichiometry can \ce{MnSnN2} accommodate? How does the cation ratio affect the unit cell? How does cation disorder affect magnetism?
To begin answering these questions, we report on our prediction, synthesis, and characterization of thin films of \ce{MnSnN2}, a new paramagnetic semiconductor.

\section{Methods} 

\subsection{Computational methods}
First-principles density functional theory (DFT) calculations were performed using the Vienna Ab initio Simulation Package (VASP).\cite{kresse1999ultrasoftVASP, shishkin2006implementationVASP} Electronic structure and band gap calculations were performed using the GW approximation \cite{hedin1965new} as described previously.\cite{lany2013NRELmatDB} Calculations were spin polarized to account for the expected magnetic moment of Mn. Computational results and details are available in the NREL Materials Database (https://materials.nrel.gov/), with ID 287033 (GW) and 287034 (DFT).

\subsection{Synthesis}
The Mn-Sn-N phase space was explored using combinatorial radiofrequency (RF) cosputtering methods.  Libraries were deposited using an AJA International ATC 2200-V sputtering chamber. The chamber base pressure was between 2 and $5\times 10^{-8}$~Torr prior to depositions.  During depositions, the pressure was maintained at $12\pm1$~mTorr with 15~sccm \ce{N2} and 5~sccm Ar. The \ce{N2} was activated by passing the gas through an electron-cyclotron resonance plasma source set at 150~W. 

To create compositional gradients within each 2" x 2" sample library, metals were cosputtered from 3" diameter Mn and Sn targets aimed towards the substrate (Kurt J. Lesker; 99.95\% and 99.998\%, respectively).  The substrate was held stationary during deposition, and the relative position of the sputtering targets (180~\degree{} relative to one another in the plane of the substrate) resulted in 1-dimensional gradients in composition across the substrate. Prior to each deposition, the targets were presputtered for 30~min with the shutters closed. Subsequently, the shutters were opened and deposition was conducted for 120~min. The sample platens were heated with lamp heaters during the depositions, with substrate temperature denoted here as $T_{\text{dep}}$. This temperature was calibrated without sputtering, and therefore provides a lower bound for the sample temperature, as the impact of sputtered elements raises the substrate temperature. Depositions were conducted between $T_{\text{dep}} = 25$~\textcelsius{} nominally (ambient conditions) and 275~\textcelsius{}. Cation composition was tuned by varying magnetron power, with Mn held at 108~W and Sn at either 36~W or 48~W. Most substrates were \textit{p}-type Si wafers with native surface oxide, although select samples were grown on Eagle Corning XG glass (EXG, optically transparent and electronically insulating) and \textit{p}-type Si with 100~nm of thermal oxide (electronically insulating) for certain property measurements as denoted in the text. 

\subsection{Characterization}
Each 2"~$\times$~2" sample library was mapped as a $4\times11$ grid following the standard combinatorial mapping workflow at the National Renewable Energy Laboratory (NREL).\cite{talley2019combigor} Experimental data used by this study have been analyzed using the COMBIgor software package\cite{talley2019combigor} and are publicly available in the high-throughput experimental materials database at https://htem.nrel.gov.\cite{zakutayev2018open_HTEM}.

High-throughput X-ray diffraction (XRD) mapping was conducted using a Bruker D8 Discover (Cu K$\alpha$ radiation) equipped with an area detector. 
High resolution synchrotron grazing incidence wide angle X-ray scattering (GIWAXS)  measurements were conducted on select samples at beamline 11-3 of the Stanford Synchrotron Radiation Lightsource, SLAC National Accelerator Laboratory with $\lambda = 0.9744$~\AA{}, a 3\textdegree{} incident angle, and spot size of 50~microns vertical $\times$ 150~microns horizontal.  
GIWAXS images were integrated with GSAS-II.\cite{toby2013gsas} 
LeBail refinements were used to extract lattice parameters for \ce{MnSnN2} in the $P6_3mc$ space group, performed with  TOPAS Academic v6.\cite{coelho2018topas}. Peak broadening was modeled via the Thompson-Cox-Hastings pseudo-Voigt “TCHZ” peak type as implemented in TOPAS. Reference patterns were generated using VESTA for visual comparisons.\cite{momma2011vesta} As the \ce{MnSnN2} phase has not been previously reported, we modeled the structure as follows: starting from the wurtzite \ce{AlN} structure,\cite{schulz1977crystalStructureAlN} we replaced \ce{Al^{3+}} with a 1:1 mixture of \ce{Mn^{2+}} and \ce{Sn^{4+}} and shifted the unit cell parameters to match the LeBail refinement for the near-stoichiometric \ce{MnSnN2} pattern (51\% Mn/(Mn+Sn)). Those refined cell parameters are $a = 3.441$~\AA{} and $c = 5.562$~\AA{} (Table \ref{tab:df_sq_table}). 

Compositional analysis was performed with X-ray fluorescence (XRF) and Rutherford Back-Scattering (RBS) methods. Metal ratios were mapped using a Fischer XDV-SDD XRF with a Rh source and a 3~mm diameter spot size. 
The measurements were performed at ambient pressure with an exposure time of 60~s for each measurement.  Nitrogen and oxygen ratios for select samples were quantified with RBS. RBS was run in a 168\textdegree{} backscattering configuration using a model 3S-MR10 RBS system from National Electrostatics Corporation with a 2 MeV \ce{He+} beam energy. Samples were measured for a total integrated charge of 160 uC. RBS spectra were modeled with the RUMP software package.\cite{barradas2008summaryRBS} Agreement on metal ratios between XRF and RBS are shown in Figure \ref{fig:RBS_merged}. A representative RBS spectrum and fit are shown in Figure \ref{fig:RBS_fit_comparison}.

Film thicknesses were measured using a Dektak profilometer and were approximately 200 to 300~nm following 2~h growth times.  Cross-sectional scanning electron microscopy (SEM) was conducted for select samples on a Hitachi S-4800 SEM operating at 3~keV accelerating voltage and 8~mm working distance to corroborate film thickness measurements and identify film morphology.

DC magnetic susceptibility was measured in a Quantum Design Magnetic Properties Measurement System (MPMS3) via vibrating sample magnetometry. The films were measured from 2~K to 300~K under applied fields from -7 to +7~T, with the applied field parallel to the plane of the film. The measured samples were 5~mm $\times$ 5~mm squares from a sample library deposited at $T_\text{dep} = 225$~\textcelsius{} on a substrate of 100~nm of \ce{SiO2} on pSi(100). Each square was characterized by XRF and synchrotron GIWAXS to ensure magnetic properties were accurately correlated with composition and structure.
To isolate the signal of the films, a bare substrate (100~nm of \ce{SiO2} on pSi(100)) was also measured and subtracted (Figure \ref{fig:backgroundSubtraction_MT} and \ref{fig:backgroundSubtraction_MH}). Additional analysis details are in the Electronic Supplementary Information.

Room temperature conductivity was measured using a custom built collinear four-point probe instrument by sweeping current between the outer two pins while measuring voltage between the inner pins (1~mm between each pin). Conventional geometric corrections were applied to convert the measured resistance into sheet resistance and then conductivity.\cite{smits1958measurement} Error bars are propagated from sheet resistance error (from I-V curve fits) and an estimated 5\% relative error in profilometry thickness measurements.

Temperature-dependent electrical conductivity was measured using a Lake Shore Cryotronics Model 8425. A 5~mm $\times$ 5~mm square was scribed out of a library deposited on EXG at approximately 200~\textcelsius{}.  Although the deposition temperature on EXG was not calibrated, XRD patterns appeared similar to those deposited on at a calibrated $T_{\text{dep}} = 225$~\textcelsius{} on silicon (i.e., XRD showed only \ce{MnSnN2} with no signs of Sn or \ce{Mn3N2} decomposition products). Indium contacts were pressed into the square near the corners of the film, and temperature-dependent sheet resistance was measured from 104~K to 298~K.
Conductivity was calculated using the profilometry-measured thickness of 260~nm. 

Spectroscopic ellipsometry was performed on a single row of a select sample library (11 points per row) using a J.A. Woollam Co. M-2000 variable angle ellipsometer. CompleteEASE software (version 6.56) was used to do the modeling. The data were modeled by fitting the imaginary part of the dielectric function with a combination of one Tauc-Lorentz, one Gaussian, and one Drude oscillator.

\section{Results and Discussion}

\subsection{Computational findings}

\begin{figure}
    \centering
    \includegraphics[width = 0.35\textwidth]{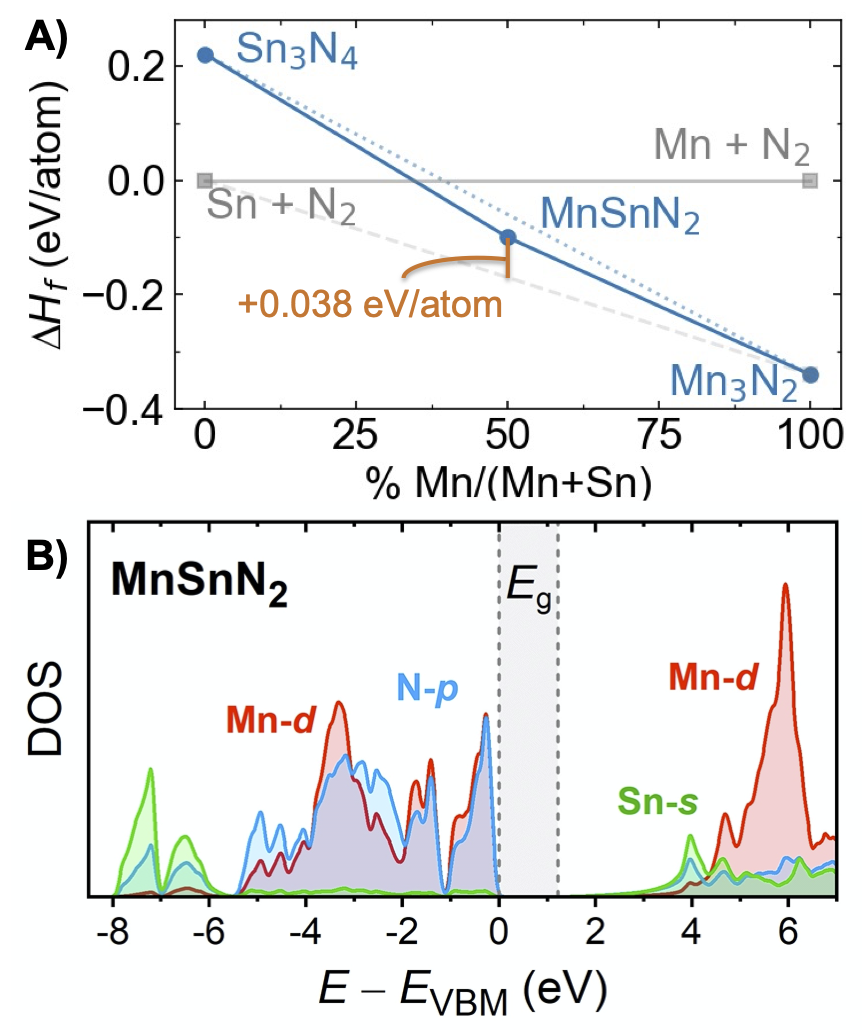}
    \caption{A) A psuedobinary phase diagram of the space from \ce{Sn3N4} to \ce{Mn3N2} at $T = 0$~K. The solid blue trace shows the hull under activated nitrogen conditions ($\mu_\mathrm{N} = +1$~eV/atom), as occurs in plasma synthesis.\cite{caskey2014thinCu3N, caskey2016synthesisSn3N4} Cation-ordered \ce{MnSnN2} is more stable than a linear combination of these two binary nitrides. B) Total density of states calculated for cation-ordered \ce{MnSnN2} in the $Pna2_1$ space group. The calculated band gap ($E_g$) is 1.22~eV, shown in grey.  N-p and Mn-d states comprise the valence band, while the conduction band consists of Sn-s and N-p states.}
    \label{fig:dft_pd_dos}
\end{figure}
To investigate the stability of a possible \ce{MnSnN2} phase, we performed density functional theory (DFT) calculations. We hypothesized that this phase would form in the wurtzite-derived structure and therefore constructed a cation-ordered \ce{MnSnN2} structure in the orthorhombic space group $Pna2_1$.  DFT calculations suggest that this cation-ordered orthorhombic \ce{MnSnN2} is slightly metastable (+0.038~eV/atom above the hull, Figure \ref{fig:dft_pd_dos}A). Therefore, a traditional bulk synthetic approach of reacting Mn and Sn under 1~atm of \ce{N2} would likely fail. However, sputtering in an activated nitrogen atmosphere (i.e., a \ce{N2} plasma) stabilizes metastable phases (e.g., \ce{Cu3N} and \ce{Sn3N4}) by increasing the chemical potential of nitrogen up to +1~eV/atom.\cite{caskey2014thinCu3N, caskey2015semiconductingSn3N4} 
Figure \ref{fig:dft_pd_dos}A shows that while \ce{MnSnN2} is metastable with respect to \ce{Sn}+\ce{N2} and \ce{Mn3N2}, the phase is stable with respect to \ce{Sn3N4} and \ce{Mn3N2}. Therefore, the nitrogen plasma environment that rendered \ce{Sn3N4} synthetically accessible should also stabilize \ce{MnSnN2}.

Our calculations also suggest that \ce{MnSnN2} should exhibit semiconducting properties and antiferromagnetic behavior. The calculated bandgap is 1.22~eV (Figure \ref{fig:dft_pd_dos}B), with a density of states (DOS) effective hole mass of $m_h^*= 5.9 m_0$. The band effective mass is $m_e^*= 0.22 m_0$. The low electron mass is remarkable for a transition metal compound (Mott insulator), but not uncommon for \ce{Mn^{2+}}.\cite{peng2015design, hautier2014does} The local magnetic moment is calculated as $\mu_\mathrm{eff} = 4.3~\mu_\mathrm{B}$ per Mn. This $\mu_\mathrm{eff}$ is less than the spin-only value expected for high-spin \ce{Mn^{2+}} ($S = \frac{5}{2}$; $\mu_\mathrm{eff} =5.9~\mu_\mathrm{B}$), indicating that some spin density resides on the nitride. The four Mn atoms per cell exhibit antiferromagnetic order in the ground state, resulting in a net moment of 0~$\mu_\mathrm{B}$ per cell.

\subsection{Synthesis and structure of \ce{MnSnN2} thin films}
\begin{figure}
\centering
\includegraphics[width=0.5\textwidth]{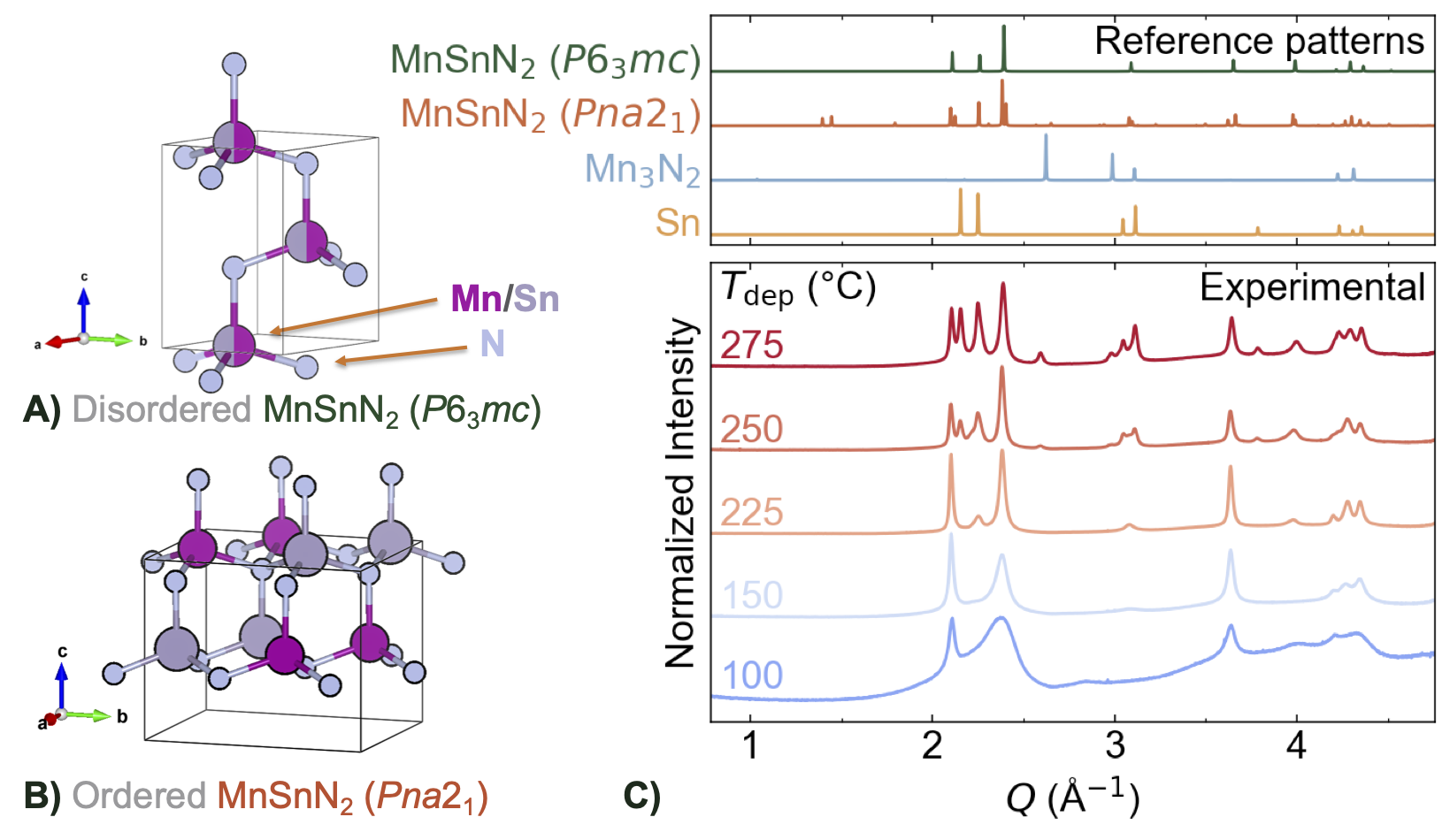}\\
\caption{A) The unit cell of \ce{MnSnN2} in the experimentally observed cation-disordered hexagonal structure ($P6_3mc$), and the B) computationally predicted ordered orthorhombic structure ($Pna2_1$). C) Synchrotron GIWAXS data of samples with Mn/(Mn+Sn)~=~50\% deposited at various temperatures ($T_{\mathrm{dep}}$) are shown in the bottom box, with reference patterns shown in the top box. The optimal deposition temperature for single phase crystalline \ce{MnSnN2} ($P6_3mc$) is $T_{\mathrm{dep}} = 225$~\textcelsius{}.  The ordered \ce{MnSnN2} ($Pna2_1$) phase is not observed.}
  \label{fgr:XRD_temp}
\end{figure}

\subsubsection{Stability and cation disorder}
We synthesized \ce{MnSnN2} films using combinatorial cosputtering and assessed the crystallinity using X-ray diffraction techniques (synchrotron GIWAXS and laboratory XRD).  Synchrotron GIWAXS of stoichiometric films shows that \ce{MnSnN2} crystallizes in a phase-pure cation-disordered wurtzite structure at $T_{\mathrm{dep}} = 225$~\textcelsius{} (Figure \ref{fgr:XRD_temp}A,C). The absence of supercell reflections (peaks near $Q = 1.5$~\AA{}$^{-1}$) of the $Pna2_1$ space group indicate that long-range ordering of cations in \ce{MnSnN2} does not occur in these samples (Figure \ref{fgr:XRD_temp}B,C). Above $T_{\mathrm{dep}} =225$~\textcelsius{}, Sn and \ce{Mn3N2} are also observed in the diffraction patterns, while below 225~\textcelsius{}, the material is poorly crystalline. These trends are true across the compositional gradient (Figure \ref{fgr:map}). This relatively low decomposition temperature ($T_{\mathrm{dep}}\sim250$~\textcelsius{}) suggests that \ce{MnSnN2} is less stable than related ternary nitrides like \ce{MgSnN2} (stable up to 500~\textcelsius{} in the same deposition chamber)\cite{greenaway2020combinatorialMgSnN2} and \ce{ZnSnN2} (stable up to 340~\textcelsius{}).\cite{fioretti2015combinatorialZnSnN2} 
Annealing experiments under flowing \ce{N2} suggest that, once deposited, \ce{MnSnN2} films are stable up to $T_{\mathrm{anneal}} =300$~\textcelsius{}, but decomposition is still observed by $T_{\mathrm{anneal}} =400$~\textcelsius{} (Figure \ref{fig:RTA}).

We conducted these annealing experiments in an attempt to induce cation ordering,\cite{schnepf2020utilizing} but saw no evidence of cation order below the decomposition temperature of $T_{\mathrm{anneal}} =400$~\textcelsius{} (Figure \ref{fig:RTA}). 
By the methods applied here, long-range cation ordering of \ce{MnSnN2} appears inaccessible. In contrast, the bulk synthesis methods employed for \ce{MnSiN2} and \ce{MnGeN2} yielded cation-ordered phases (space group $Pna2_1$).\cite{esmaeilzadeh2006crystalMnSiN2, wintenberger1972etudeMnGeN2}  The prior report on thin films of \ce{MnGeN2} did not assess cation ordering via diffraction.\cite{liu2012structural_MBE_MnGeN2}

The lack of long-range cation ordering exhibited by \ce{MnSnN2} is common in sputtered films. \ce{MgSnN2} and \ce{ZnSnN2} thin films are both cation disordered,\cite{greenaway2020combinatorialMgSnN2, fioretti2015combinatorialZnSnN2} and \ce{ZnGeN2} forms as a cation-disordered thin film but crystallizes in the ordered $Pna2_1$ structure when synthesized by bulk ammonolysis.\cite{melamed2020combinatorialZnGeN2, blanton2017characterizationOrderZnGeN2} The high effective temperature of the sputtering synthesis technique, along with the entropic benefit of disorder, can stabilize a disordered structure by approximately 0.1~eV/atom relative to the DFT-predicted ordered structures.\cite{woods2022role} II-Sn-\ce{N2} phases in particular have a tendency for disorder because \ce{Sn^{4+}} is closer in ionic radius (0.55~\AA{}) to the divalent cations---\ce{Mn^{2+}} (0.66~\AA{}), Mg (0.57~\AA{}), and Zn (0.60~\AA{})---than the other main group elements---\ce{Ge^{4+}} (0.39~\AA{}) and \ce{Si^{4+}} (0.26~\AA{}).\cite{greenaway2020combinatorialMgSnN2, khan2020reviewZnSnN2, shannon1969effectiveradii} This size similarity leads to cation disorder even in bulk \ce{MgSnN2}.\cite{kawamura2020synthesisMgSnN2} 
While short-range ordering is possible even with the absence of a supercell reflection and may impact properties,\cite{schnepf2020utilizing, quayle2015chargeNeutrality_II_IV_V2, melamed2022shortrangeOrderingZnSnN2} such analysis is beyond the scope of this study.

\subsubsection{Compositional and structural trends}

\begin{figure*}
\centering
\includegraphics[width=0.9\textwidth]{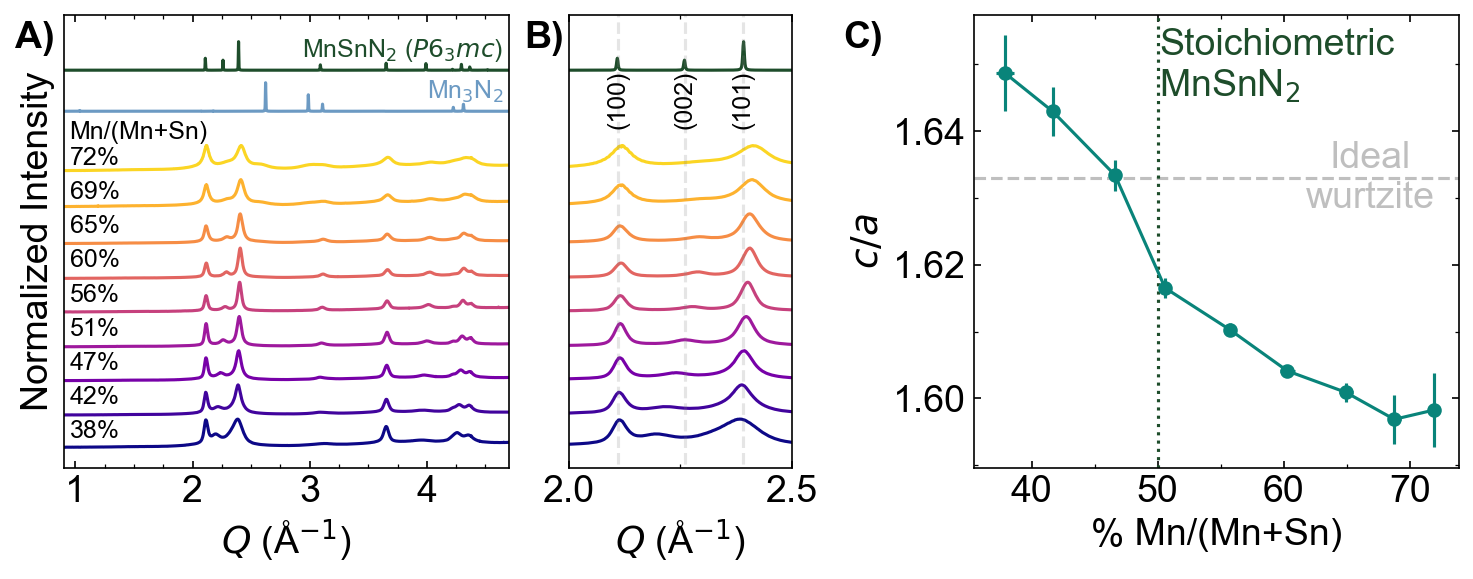}\\
  \caption{A) Synchrotron GIWAXS data of select samples from a combinatorial library row deposited at $T_{\text{dep}} = 225$~\textcelsius{} shows that reflections can be indexed to the $P6_3mc$ space group, indicative of a cation-disordered wurtzite structure. B) Magnified $Q$ range of the diffraction patterns, showing the shift in the (002) reflection with changing Mn/(Mn+Sn). C) The \textit{c} axis contracts relative to the \textit{a} axis with increasing Mn content, which causes the $c/a$ ratio to decrease. The vertical dotted line indicates the composition of stoichiometric \ce{MnSnN2} and the horizontal dashed line indicates the ideal \textit{c/a} ratio of the wurtzite structure ($\sqrt{8}/3$).  Fits to the data are shown in Figure \ref{fgr:XRD_fits_SI}. 
  }
  \label{fgr:vegard}
\end{figure*}

Despite this low thermal stability, the wurtzite structure of \ce{MnSnN2} accommodates a wide range of cation off-stoichiometry ($18\% < $ Mn/(Mn+Sn) $ < 65\%$) when deposited at the optimal temperature of $T_{\text{dep}} = 225$~\textcelsius{}.
Within the composition range of Mn/(Mn+Sn) = 38\% to 65\%, all reflections in the GIWAXS patterns can be indexed to $P6_3mc$ symmetry (Figure \ref{fgr:vegard}). 
In laboratory XRD, we observe \ce{MnSnN2} down to Mn/(Mn+Sn) = 18\% (the Mn-poor limit of the library, Figure \ref{fig:Mn_poor_XRD}). However, this may not be the true lower compositional limit of this phase. 
In contrast, we have determined that the upper limit of Mn content in \ce{MnSnN2} by diffraction is Mn/(Mn+Sn) $\approx 65\%$.
At high Mn content ($>65$\%), the lattice parameters of \ce{MnSnN2} plateau and the uncertainty in the refinement grows (Figures \ref{fgr:vegard}C and \ref{fig:refinement_parameters}).
Broad peaks that index to \ce{Mn3N2} also appear, indicative of a two-phase region.
The only other known Mn-Sn-N ternary compound is the antiperovskite \ce{Mn3SnN} (Mn/(Mn+Sn) = 75\%),\cite{nardin1972retudeMn3SnN, mekata1962magneticMn3SnN} which is more Mn-rich than our depositions and does not appear in our diffraction patterns.

This wide range of off-stoichiometry can make naming ternary nitrides difficult, as noted by Woods-Robinson, et al.\cite{woods2022role} 
A precise formula would be \ce{Mn$_x$Sn$_{1-x}$N$_{1-y}$O$_y$}, where $x$ is the cation fraction of Mn/(Mn+Sn) and $y$ is the amount of oxygen substitution for nitrogen (estimated to be $y\approx0.2$ by RBS, Figure \ref{fig:RBS_merged}B). The oxygen likely originates from oxide contamination of the Mn sputter target, and either charge balances with \ce{Mn^{2+}} via (\ce{MnSnN2})$_{1-x}$(MnO)$_x$ or with \ce{Mn^{3+}} via \ce{Mn$_{1+x}$Sn$_{1-x}$N$_{2-x}$O$_x$}.
For simplicity, we use ``\ce{MnSnN2}'' to refer to the cation-disordered wurtzite-like phase across the whole composition range and specify a cation fraction as a percentage of Mn/(Mn+Sn) where relevant (i.e., stoichiometric \ce{MnSnN2} has Mn/(Mn+Sn) = 50\%).

\begin{figure}
    \centering
    \includegraphics[width = 0.35\textwidth]{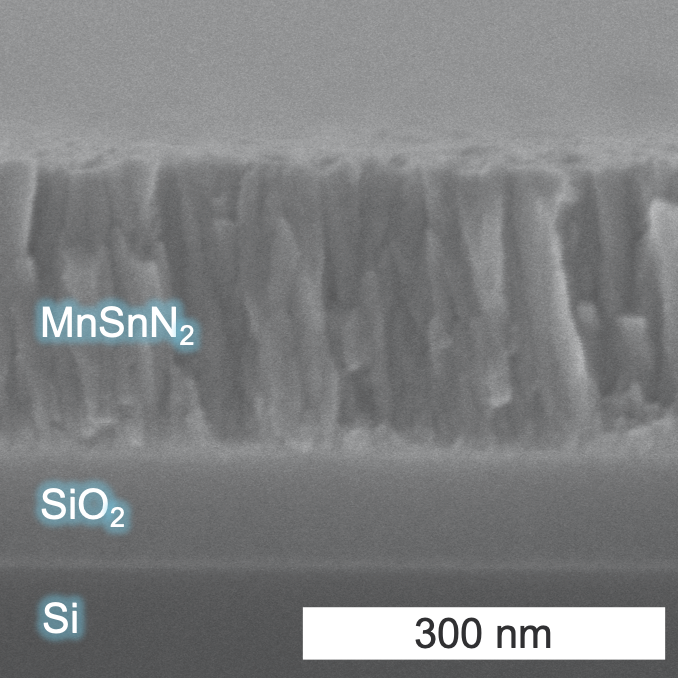}
    \caption{Cross-sectional SEM micrograph showing columnar grains typical of sputtered films. This representative image shows a 50\% Mn/(Mn+Sn) film deposited on a silicon substrate with 100~nm of \ce{SiO2}. The scale bar is 300~nm.}
    \label{fig:SEM}
\end{figure}

The \textit{c/a} ratio of \ce{MnSnN2} varies with composition, but deviates only slightly from the ideal wurtzite value of 1.633 (Figure \ref{fgr:vegard}C).
For stoichiometric \ce{MnSnN2}, \textit{c/a} =  $5.562$~\AA{} / $3.441$~\AA{} $= 1.616$, which---although smaller than the ideal wurtzite---is larger than \ce{MgSnN2} (1.598).\cite{greenaway2020combinatorialMgSnN2}
The change in \textit{c/a} is driven primarily by a change in the \textit{c} lattice parameter while the \textit{a} lattice parameter remains largely unchanged (Figure \ref{fig:refinement_parameters}A), suggesting that the tetrahedra may be distorting along the \textit{c} axis. 
That shift may be caused by metal vacancies, a change in electron count of the metals, oxygen incorporation (observed by RBS, Figure \ref{fig:RBS_merged}B), or some combination thereof as the system shifts to maintain charge neutrality across this compositional range. 
However, since atomic positions cannot be resolved from the synchrotron GIWAXS data sets, we cannot more precisely identify the cause of this distortion.

To probe the microstructure of the film, we turn to electron microscopy and diffraction analysis. SEM shows columnar grain morphology that is typical for sputtered films (Figure \ref{fig:SEM}).  Diffraction images display some signs of texturing (Figure \ref{fgr:representive_2D_GIWAXS}). The (002) reflection appears towards the edges of the integration window, suggesting the \{002\} planes are oriented normal to the plane of the substrate. In contrast, the (100) reflection appears strongly in the center axis of the frame, 90~\textdegree{} relative to the (002) reflection, and diminishes in intensity towards the edge of the integration window, suggesting preferential growth along the [100] direction (Figure \ref{fgr:vegard}B). 
This finding stands in contrast to typical wurtzite films, which tend to grow with (002) preferential orientation.\cite{fay2005lowZnO, fioretti2015combinatorialZnSnN2} 

\subsection{Magnetic properties}
Given the presence of tetrahedrally coordinated \ce{Mn^{2+}} (expected: $S=\frac{5}{2}$ and $\mu_{\mathrm{eff}} = $5.9~$\mu_\mathrm{B}$) cations in stoichiometric \ce{MnSnN2}, as well as the possible presence of \ce{Mn^{3+}} (expected: $S=2$ and $\mu_{\mathrm{eff}} = $4.9~$\mu_\mathrm{B}$) in Mn-rich compositions, DC magnetic susceptibility was employed to investigate the magnetic response of several thin films with a range of compositions 42\% $\leq$ Mn/(Mn+Sn) $\leq$ 56\%. We cannot independently determine the film mass, so susceptibility ($\chi$) was calculated from the raw moment data using the same estimated mass (0.04~mg) for each sample. Uncertainty in film thickness, composition, and porosity may introduce substantial error in the calculated mass of the films, and by extension, the absolute value of $\chi$.  In addition, each measured sample contains a compositional gradient of approximately 4\% Mn/(Mn+Sn). Despite these caveats, errors from sample to sample are likely systematic, allowing us to identify trends in magnetism related to composition. Additional details and discussion are in the Electronic Supplementary Information (Section 3).

\begin{figure*}
\centering
\includegraphics[width = 0.9\textwidth]{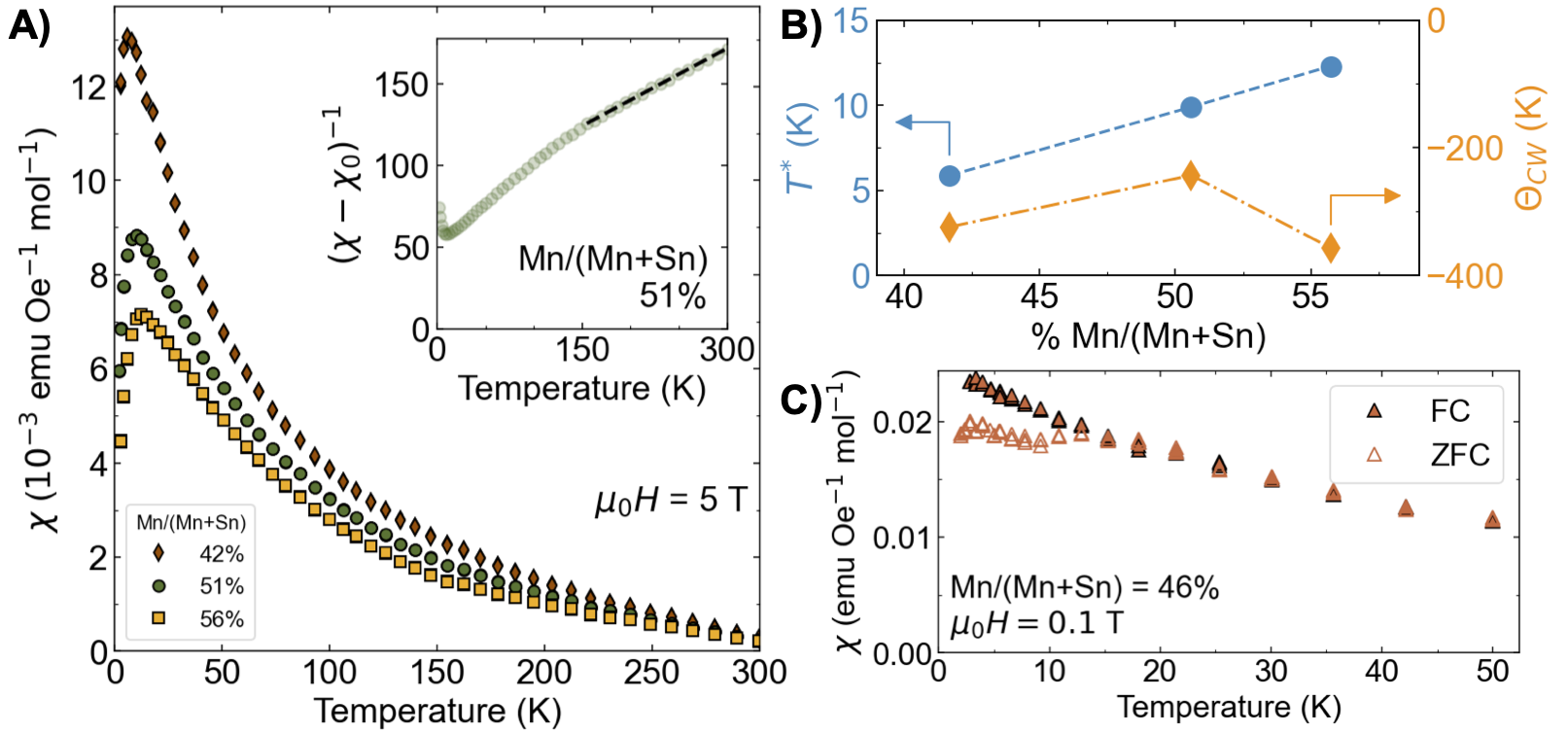}
  \caption{A) Zero-field-cooled (ZFC) DC magnetic susceptibility ($\chi$) as a function of temperature measured in an applied field of $\mu_0H = 5$ T, show a peak in $\chi$ at $T^\mathrm{*} \approx 10$~K across a range of Mn contents.  The inset shows  $\chi^{-1}$ as a function of temperature with a diamagnetic correction $\chi_0 = -5.5\times10^{-3}$ emu Oe$^{-1}$ mol$^{-1}$ for 51\% Mn/(Mn+Sn). The dashed trace shows the Curie-Weiss fit from 150~K to 300~K. B) Extracted transition temperatures ($T^\mathrm{*}$, left axis) and Weiss temperatures ($\Theta_{CW}$, right axis) as a function of increasing Mn content. C) ZFC and field-cooled (FC) DC magnetic susceptibility measurements at lower applied field of $\mu_0H = 0.1$~T show bifurcation below the transition temperature. } 
  \label{fgr:magnetism_chi_neel}
\end{figure*}

As shown in Figure \ref{fgr:magnetism_chi_neel}A, zero-field-cooled (ZFC) temperature-dependent susceptibility measurements performed at high applied field exhibit peaks near $10$~K, indicating a magnetic transition ($T^\mathrm{*}$).  
The $T^\mathrm{*}$, extracted as the temperature where $\chi$ reaches a maximum, increases slightly with increasing Mn content, from $T^\mathrm{*} \approx 6$~K at 42\% Mn to $T^\mathrm{*} \approx 12$~K at 56\% Mn (Figure \ref{fgr:magnetism_chi_neel}B). 
Curie-Weiss fits in the paramagnetic region 150--300~K (see inset of Figure \ref{fgr:magnetism_chi_neel}A and Figure \ref{fig:CW_fits}) show large, negative Weiss temperatures ($\Theta_{CW}$, Figure \ref{fgr:magnetism_chi_neel}B), consistent with overall antiferromagnetic (AFM) correlations. However, the uncertainties in background subtraction and in the calculation of $\chi$ inhibit meaningful extraction of Curie constants and effective moments from these fits (see Electronic Supplementary Information Section 3). 
The magnitude of $\chi$ decreases with increasing Mn fraction, which would be consistent with \ce{Mn^{3+}} replacing \ce{Mn^{2+}}. However, we are not able to detect a significant amount of \ce{Mn^{3+}} in XANES measurements (Figure \ref{fig:xanes}). This apparent shift in $\chi$ may therefore be an artifact of the uncertainty in film mass. 
Measurements in lower applied field ($\mu_0H = 0.1$~T; Figure \ref{fgr:magnetism_chi_neel}C) show clear bifurcation in ZFC and field-cooled (FC) susceptibility  below the transition temperature, which is consistent with either  long-range AFM order or a spin-glass ground state.

\begin{figure}
    \centering
    \includegraphics[width = 0.35\textwidth]{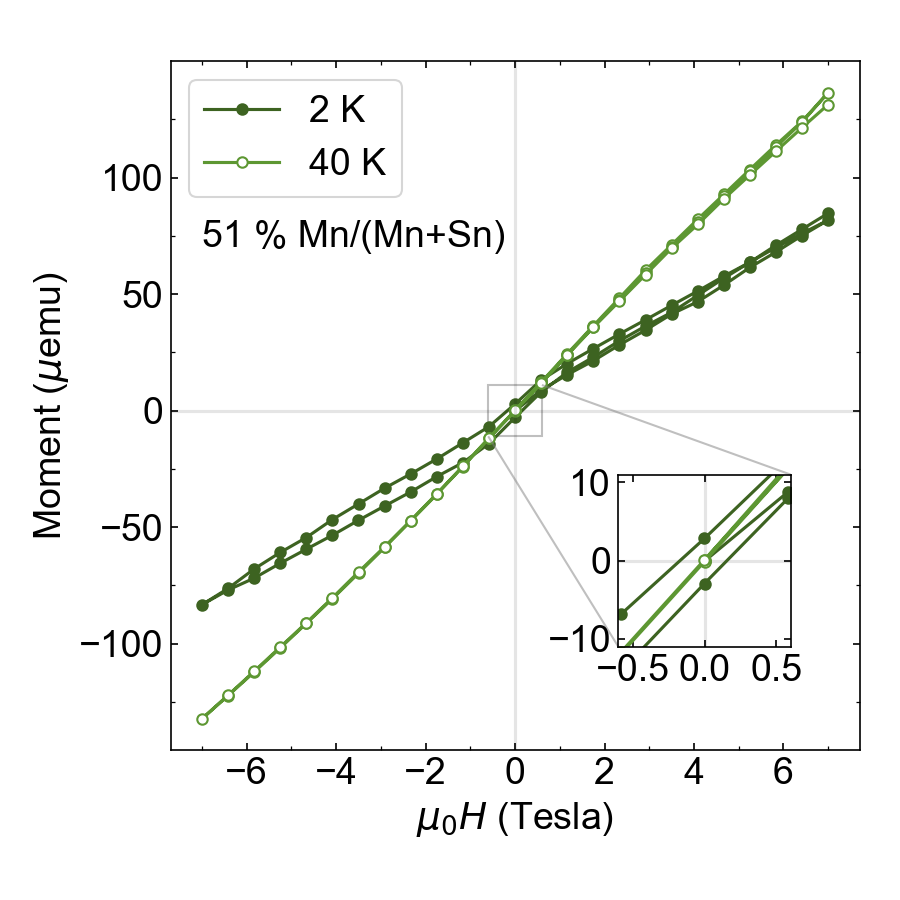}
    \caption{Background-subtracted isothermal magnetization measurements of near-stoichiometric \ce{MnSnN2} 
    at $T=2$~K (solid circles) and $T=40$~K (open circles).
    The inset details the small net ferromagnetic moment at $T=2$~K and the absence of such net moment at $T=40$~K.}
    \label{fig:Magnetism_MH_2K_40K}
\end{figure}

The magnetic moment measured as a function of applied field at $T=2$~K (below the transition at $T^\mathrm{*} \approx 10$~K) is consistent with AFM correlations, shown for Mn/(Mn+Sn) = 51\% in Figure \ref{fig:Magnetism_MH_2K_40K} and for the remaining samples in Figure \ref{fgr:magnetism_MH}.
In the $T=2$~K data of all samples measured, there is small but observable hysteresis with a minute net moment at zero field (inset of Figure \ref{fig:Magnetism_MH_2K_40K}). 
This net moment may originate from spin canting, uncompensated moments from defects, the Dzyaloshinsky-Moriya interaction, or slow magnetic relaxation of a spin glass state. Similar net moments were observed in thin films of \ce{MnGeN2} and were attributed to cation site disorder.\cite{liu2012structural_MBE_MnGeN2}
Above $T^\mathrm{*}$, (e.g., at $T=40$~K), the linear behavior is consistent with paramagnetism. The traces do not plateau at high fields, indicating that the magnetic moment is not saturated at $\mu_0H=\pm7$~T.

Overall, the susceptibility measurements and Curie-Weiss fit results point to clear, strong AFM correlations in \ce{MnSnN2} with a transition at approximately $T^\mathrm{*}\approx 10$~K. This transition shifts as a function of Mn/(Mn+Sn). We are unable to determine whether the ground state below this transition temperature is long-range AFM order or spin glass.\cite{mugiraneza2022tutorial_magnetism} If the ground state is an ordered AFM, it would likely be complex, especially given the cation disorder and compositional gradients in these samples. The small net moment observed at $T < T^\mathrm{*}$ may imply the presence of spin canting, Dzyaloshinsky-Moriya interactions, or uncompensated residual spins, as mentioned above, and the ZFC-FC splitting suggests a relaxation process. However, the Mn content in these samples (41--56\%) is well above the percolation threshold (20\% for this hexagonal close packed metal sublattice),\cite{van1997percolation} so the AFM correlations likely extend throughout the crystalline domains. 
To probe the possibility of spin glass behavior, we performed AC susceptibility measurements, but the signal was below the instrumental sensitivity limit. 

The low-temperature magnetic transition ($T^\mathrm{*} \approx 10$~K) of the cation-disordered \ce{MnSnN2} thin films examined here contrasts starkly with the high-temperature AFM ordering of cation-ordered \ce{MnGeN2} ($T_N = 448$~K) and \ce{MnSiN2} ($T_N = 490$~K) synthesized as bulk single crystals.\cite{wintenberger1972etudeMnGeN2, esmaeilzadeh2006crystalMnSiN2} 
To explain these differences, we posit that cation ordering is an important factor in determining the energy scale of magnetic order in this family of wurtzite ternary nitrides. 
That influence may be exerted via magnetic frustration, if the ground state is AFM order: the tetrahedrally-based lattice of \ce{Mn^{2+}}, together with the large ratio of the Weiss temperatures ($\Theta_{CW}\approx-200$~K) and the low transition temperatures ($T^\mathrm{*}\approx 10$~K) of these films, suggest the presence of a high degree of magnetic frustration.\cite{mugiraneza2022tutorial_magnetism, Ramirez1994}  
This is not unexpected, as triangular-based lattices are one of the most common hosts of geometric frustration, and frustration also often arises from cation disorder.\cite{Ney2016,Simonov2020} 
In addition, frustration was observed even in the cation-ordered \ce{MnSiN2}.\cite{esmaeilzadeh2006crystalMnSiN2} 
Should bulk powders or single crystals of \ce{MnSnN2} be synthesized, comparing their magnetic properties and defect chemistry with the thin film behavior observed here would be illuminating.

\subsection{Optoelectronic properties}
\begin{figure*}[ht!]
\centering
\includegraphics[width=0.9\textwidth]{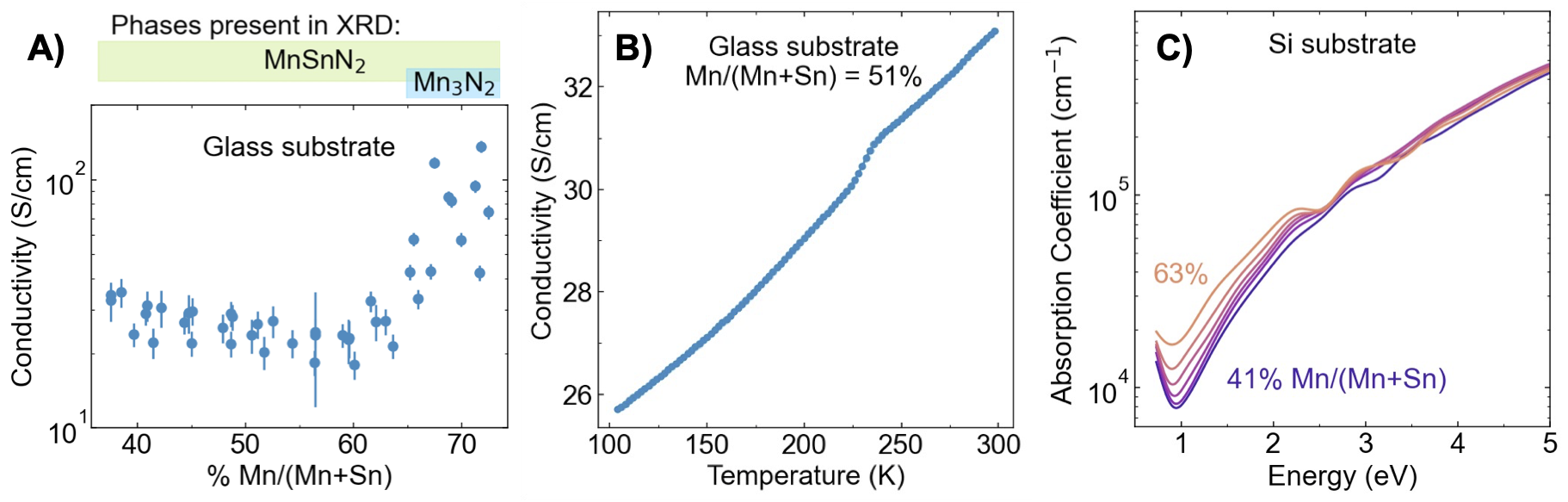}
  \caption{A) Room temperature conductivity measurements performed in a collinear four-point probe configuration show that conductivity initially decreases slightly with increasing Mn content, then increases rapidly above 60\% Mn/(Mn+Sn). B) Conductivity increases as a function of temperature, confirming the semiconducting nature of \ce{MnSnN2} films. The anomaly in slope near 250~K is an instrumental artifact. C) Optical ellipsometry measurements indicate a gradual absorption onset, with minima in the absorption coefficients near 1~eV. 
  }
  \label{fgr:optoelectronics}
\end{figure*}
Room temperature conductivity measurements show that the \ce{MnSnN2} films exhibit conductivity on the order of 20~S/cm (Figure \ref{fgr:optoelectronics}A).  Conductivity decreases slightly as Mn content increases from 38\% to 60\% Mn/(Mn+Sn) but then increases as a metallic secondary phase (\ce{Mn3N2}) grows in. This increase in conductivity for Mn/(Mn+Sn) $\geq60$\% suggests that the true single-phase region may have a lower Mn limit ($\sim$60\%  maximum) than detected by GIWAXS ($\sim$65\%  maximum). 
For the Mn/(Mn+Sn) $<60$\% region, conductivity is likely tuned by carrier concentration. 
Prior work on \ce{ZnSnN2} demonstrated that a similar trend (decreasing conductivity with increasing Zn) was correlated with a decrease in carrier concentration (while mobility was unaffected).\cite{fioretti2015combinatorialZnSnN2} 
In both cases (\ce{MnSnN2} and \ce{ZnSnN2}), the conductivity reaches a minimum in the Sn-poor region (i.e., Zn or Mn $\approx$60\%), as the excess Zn or Mn compensate for oxygen substitution on nitrogen sites. At Mn/(Mn+Sn) = 60\%, the charge-balanced chemical formula is  \ce{Mn_{0.6}Sn_{0.4}N_{0.8}O_{0.2}} (assuming \ce{Mn^{2+}} and \ce{Sn^{4+}}), consistent with the oxygen content identified by RBS (Figure \ref{fig:RBS_merged}B). 

Temperature-dependent conductivity measurements performed in Van der Pauw geometry on films grown on insulating EXG glass show that the conductivity increases with increasing temperature, confirming semiconducting behavior in \ce{MnSnN2} (Figure \ref{fgr:optoelectronics}B). Increasing conductivity with increasing temperature suggests that mobile carriers are thermally activated (i.e., a bandgap is present in the material). However, the conductivity is only weakly dependent on temperature, with the conductivity at 298~K only 30\% greater than the conductivity at 104~K.  

Ellipsometry measurements and modeling show that \ce{MnSnN2} has a gradual absorption onset near 1~eV (Figure \ref{fgr:optoelectronics}C). Overall, the absorptivity increases slightly with increasing Mn content, although the absorption onset stays relatively constant.  This absorption onset may be attributable to a bandgap, as the bandgap for this material is expected to be lower than the experimentally reported bandgaps in cation-ordered \ce{MnSiN2} (3.5~eV) and \ce{MnGeN2} (2.5~eV).\cite{hausler2018ammonothermal} The DFT-calculated bandgap for cation-ordered \ce{MnSnN2} is 1.22~eV (Figure \ref{fig:dft_pd_dos}B), and the cation-disorder of the material synthesized here tends to lower the bandgap relative to cation-ordered analogues.\cite{schnepf2020utilizing}
However, \ce{Mn^{2+}} \textit{d}-\textit{d} transitions are also known to occur in this energy region,\cite{esmaeilzadeh2006crystalMnSiN2} and unfortunately, this technique cannot distinguish between single-ion electronic transitions and bandgap excitations. 
We also attempted UV-vis measurements of films deposited on transparent substrates to verify the optical absorption onset, but the onset was below the limit of the instrument (Figure \ref{fgr:uvvis}). Regardless, the optoelectronic and magnetic behavior of this new compound present opportunities for further study.

\section{Conclusions}
We have synthesized thin films of a new ternary nitride material, \ce{MnSnN2}, using combinatorial co-sputtering. The material crystallizes in a cation-disordered wurtzite-like structure common for II-IV-\ce{N2} semiconductors with lattice parameters $a=3.441$~\AA{} and $c = 5.562$~\AA{} for the stoichiometric material. \ce{MnSnN2} exhibits semiconducting properties, including a bandgap near 1~eV---the lowest bandgap yet observed in the \ce{Mn\textit{M}N2} wurtzite family.\cite{hausler2018ammonothermal}  We observe a magnetic transition in cation-disordered \ce{MnSnN2} at low temperature ($T^\mathrm{*}\approx$10~K) consistent with AFM correlations, which contrasts dramatically with the high AFM ordering temperatures of cation-ordered \ce{MnGeN2} and \ce{MnSiN2} ($T_N = 448$~K and 490~K, respectively).\cite{esmaeilzadeh2006crystalMnSiN2, wintenberger1972etudeMnGeN2} These discoveries suggest that future work could use composition and cation disorder in the Mn-$M$-\ce{N2} ($M$ = Si, Ge, Sn) system to tune both semiconducting and magnetic properties. 

\section*{Author Contributions}
J.S.A. and A.C.T. conceived of the project. J.S.A. and C.L.R. conducted depositions, XRF, and XRD measurements. C.L.R. and R.W.S conducted magnetism and electronic conductivity measurements. R.R.S. and C.L.M. conducted ellipsometry measurements. L.T.S. and C.L.R. conducted synchrotron GIWAXS measurements. K.N.H. conducted RBS analysis. J.S.M. conducted microscopy. S.-J.L. conducted XAS measurements. S.L. conducted DFT calculations. C.L.M. and A.L.G. aided in synthesis and maintenance of the deposition chamber. C.L.R. and R.W.S. wrote the majority of the manuscript with guidance from S.R.B., J.R.N., A.C.T., and J.S.A as well as feedback from all other authors.

\section*{Conflicts of interest}
There are no conflicts of interest to declare

\section*{Acknowledgements}
This work was authored in part by the National Renewable Energy Laboratory (NREL), operated by Alliance for Sustainable Energy, LLC, for the U.S. Department of Energy (DOE) under Contract No. DE-AC36-08GO28308. Primary funding was provided by the U.S. Department of Energy, Office of Science, Basic Energy Sciences, Materials Sciences and Engineering Division. C.L.R. acknowledges support from the DOE Science Graduate Research Program (SCGSR). R.W.S. acknowledges support from the Director’s Fellowship within NREL’s Laboratory Directed Research and Development program. Use of the Stanford Synchrotron Radiation Lightsource, SLAC National Accelerator Laboratory, is supported by the U.S. Department of Energy, Office of Science, Office of Basic Energy Sciences under Contract No. DE-AC02-76SF00515. We thank the Analytical Resources Core at Colorado State University for instrument access and training. We thank Charles Titus and Andriy Zakutayev for helpful conversations. The views expressed in the article do not necessarily represent the views of the DOE or the U.S. Government.

\bibliography{MnSnN2_synth.bib}
\bibliographystyle{rsc} 

\clearpage
\onecolumn 

\renewcommand{\thefigure}{S\arabic{figure}}
\setcounter{figure}{0}
\renewcommand{\thesection}{S\arabic{section}}
\setcounter{section}{0}
\renewcommand{\theequation}{S\arabic{equation}}
\renewcommand{\thetable}{S\arabic{table}}
\renewcommand{\thepage}{S\arabic{page}}
\setcounter{page}{1}

\section*{\dag ELECTRONIC SUPPLEMENTARY INFORMATION}

\tableofcontents

\section{Structural characterization}
\begin{figure}[hb]     
\centering
\includegraphics[width=0.8\textwidth]{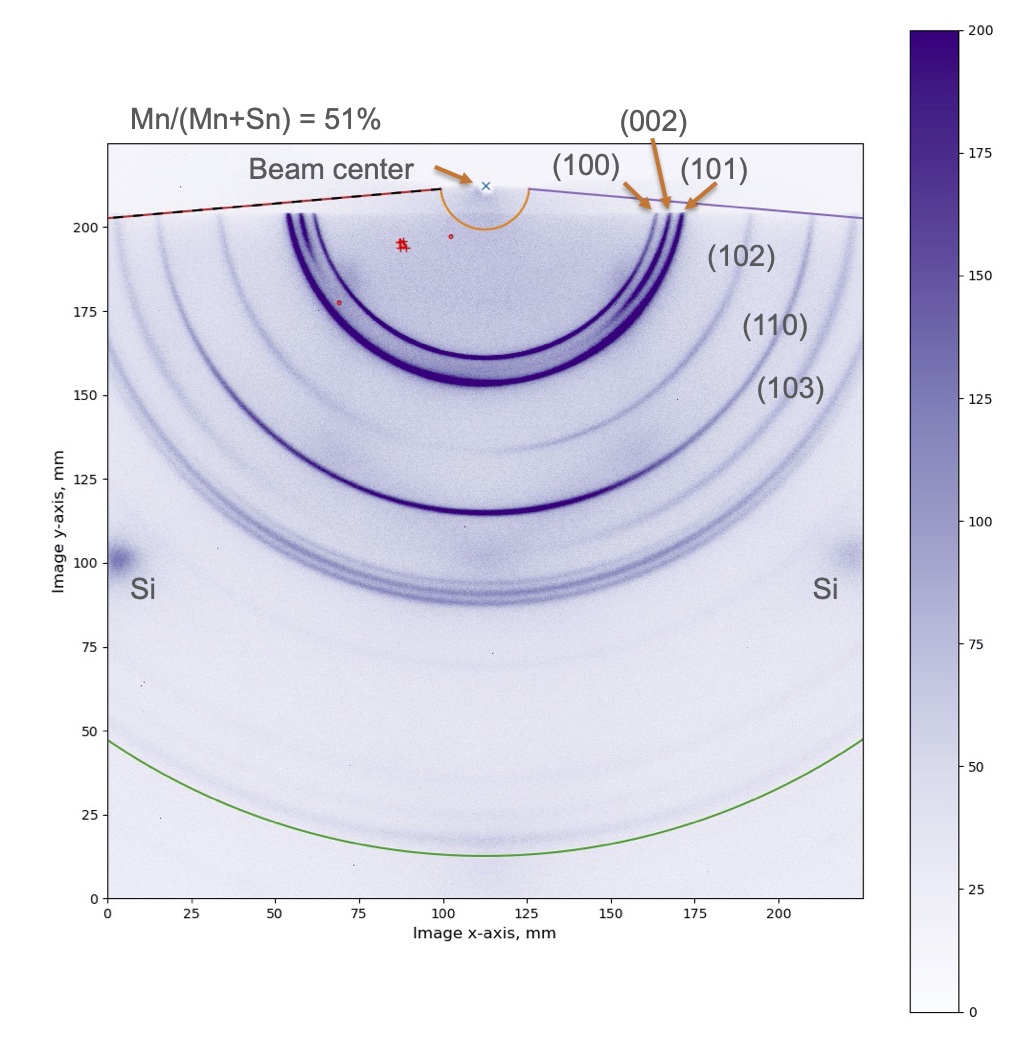}\\
  \caption{A representative 2-D diffraction image from a GIWAXS measurement at SSRL beamline 11-3. Hot pixels are masked in red to avoid background contribution to the integrated diffraction pattern. The first six reflections of \ce{MnSnN2} are also labeled. Strong spots from the Si substrate (331) planes are visible near the edge of the frame.}
  \label{fgr:representive_2D_GIWAXS}
\end{figure}

\begin{figure}
\centering
\includegraphics[width=0.5\textwidth]{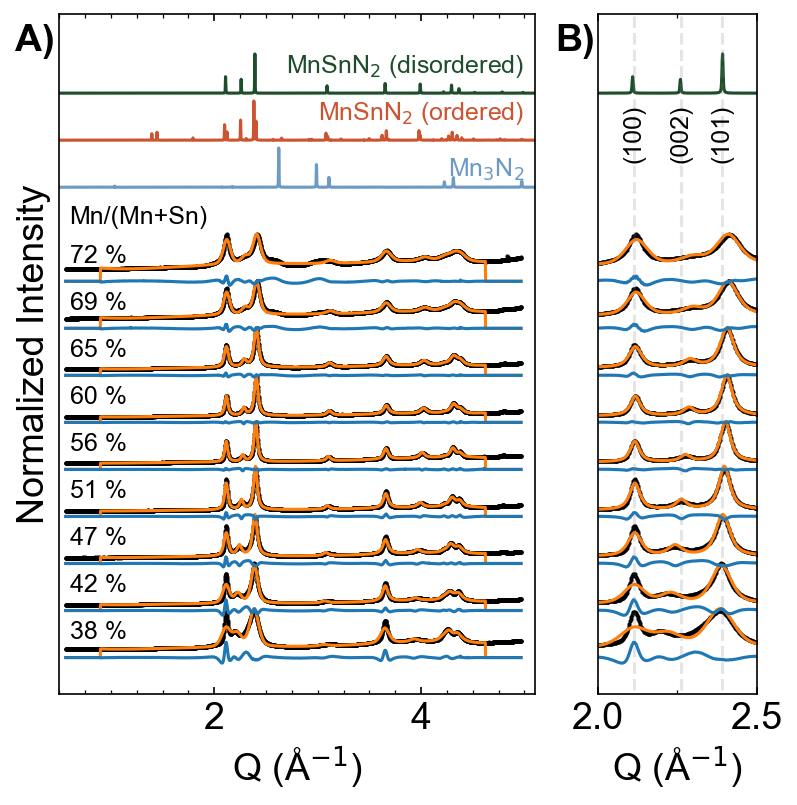}\\
  \caption{A) Full patterns of the GIWAXS data presented in Figure \ref{fgr:vegard} of the main text. 
  VESTA-generated reference patterns are shown at the top.
  The main reflections match the $P6_3mc$ space group of the cation-disordered wurtzite structure type \ce{MnSnN2}. Evidence of cation ordering is not observed in diffraction. At Mn-rich compositions ($\geq 69$~\%), broad peaks corresponding to \ce{Mn3N2} appear.
  LeBail refinements were conducted on these patterns to extract lattice parameters of the wurtzite phase \ce{MnSnN2}. Black points are data, orange traces are fits, and blue traces are difference curves.
 B) Focused $Q$-range of fits to the first three wurtzite reflections, corresponding to Figure \ref{fgr:vegard}B of the main text. Although the peak broadness is poorly captured owing to anisotropic broadening, the peak positions are relatively well described by the fit.
  }
  \label{fgr:XRD_fits_SI}
\end{figure}

\begin{figure*}
    \centering
    \includegraphics[width = 0.9\textwidth]{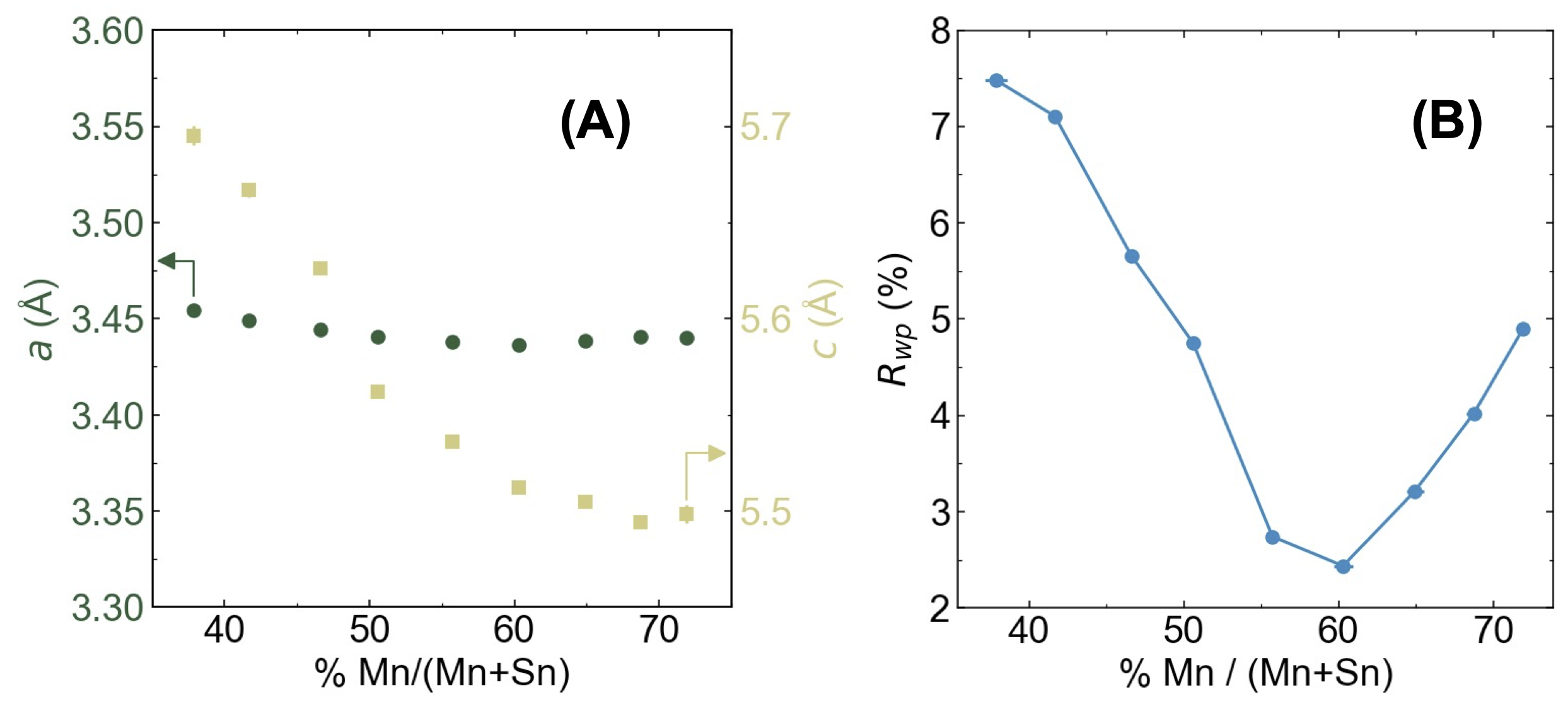}
    \caption{A) Extracted lattice parameters $a$ (green circles) and $c$ (gold squares) from LeBail refinements used to calculate the $c/a$ ratio in Figure \ref{fgr:vegard}C. Error bars are within the size of the markers.
    B) Residuals from the LeBail refinements presented in Figure \ref{fgr:XRD_fits_SI}. The modeled $P6_3mc$ phase fits best with the slightly Mn-rich composition. Anisotropic peak broadening worsens the fit in the Mn-poor region. A poorly crystalline \ce{Mn3N2} secondary phase, which was not included in the model, worsens the fit in the Mn-rich region.}
    \label{fig:refinement_parameters}
\end{figure*}

Synchrotron grazing incidence wide angle X-ray scattering (GIWAXS) reveals the polycrystalline nature of the films deposited on silicon with a 100 nm surface layer of \ce{SiO2} (Figure \ref{fgr:representive_2D_GIWAXS}).  Hot pixels were manually masked when identified to avoid spurious peaks in the integrated patterns. Broad and weak reflection spots are visible and attributable to silicon, but these spots were not masked owing to significant overlap with the powder rings. These spots contribute to elevated background scattering in the integrated diffraction patterns. Full LeBail fits for the patterns shown in Figure \ref{fgr:vegard} are shown in Figure \ref{fgr:XRD_fits_SI}.  Figure \ref{fig:refinement_parameters}A shows the extracted hexagonal lattice parameters and Figure \ref{fig:refinement_parameters}B show the fit residuals.

Fitting residuals were minimized for the patterns with slight excess Mn, with worsening fits in the Mn-poor and Mn-rich regions (Figure \ref{fig:refinement_parameters}B). 
Fit residuals also increase in the Mn-rich region, here owing to the growth of a secondary \ce{Mn3N2} phase that is visible as broad peaks but was not modeled. Some of this broadening may also be due to scattering from the silicon substrate, which may vary from sample to sample depending on the exact orientation of the substrate. Lastly, some degree of texturing is observed (Figure \ref{fgr:representive_2D_GIWAXS}), which affects the relative heights of integrated peaks.

\begin{figure*}
\centering
\includegraphics[width= 0.9\textwidth]{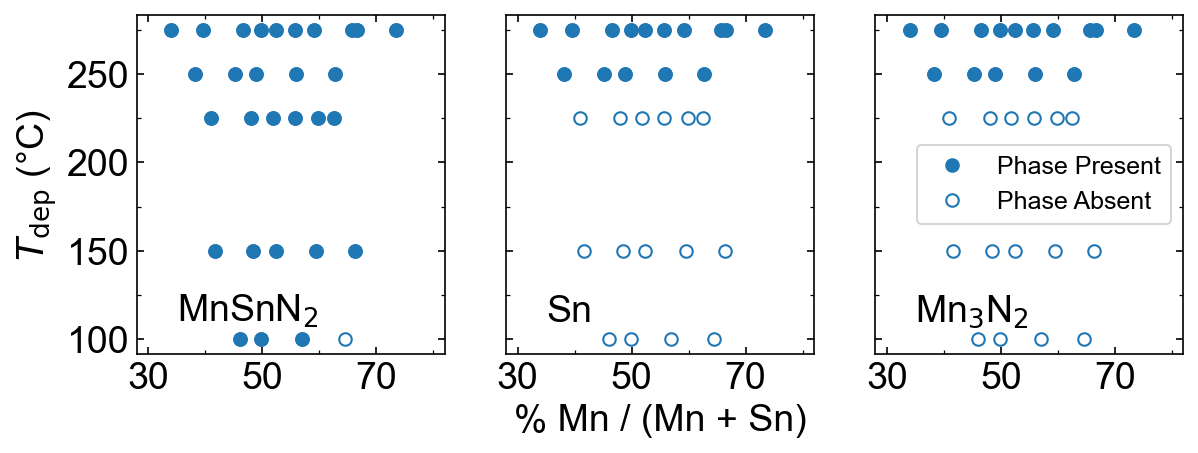}\\
  \caption{GIWAXS patterns were collected on select points from a variety of libraries. Manual inspection revealed the predominance of a single phase, \ce{MnSnN2} up to 225~\textcelsius{}. However, at higher temperatures, the \ce{MnSnN2} phase begins decomposing into \ce{Mn3N2} and \ce{Sn}. The coexistence of all three suggests non-equilibrium conditions.}
  \label{fgr:map}
\end{figure*}

\begin{figure}     \centering
\includegraphics[width=0.5\textwidth]{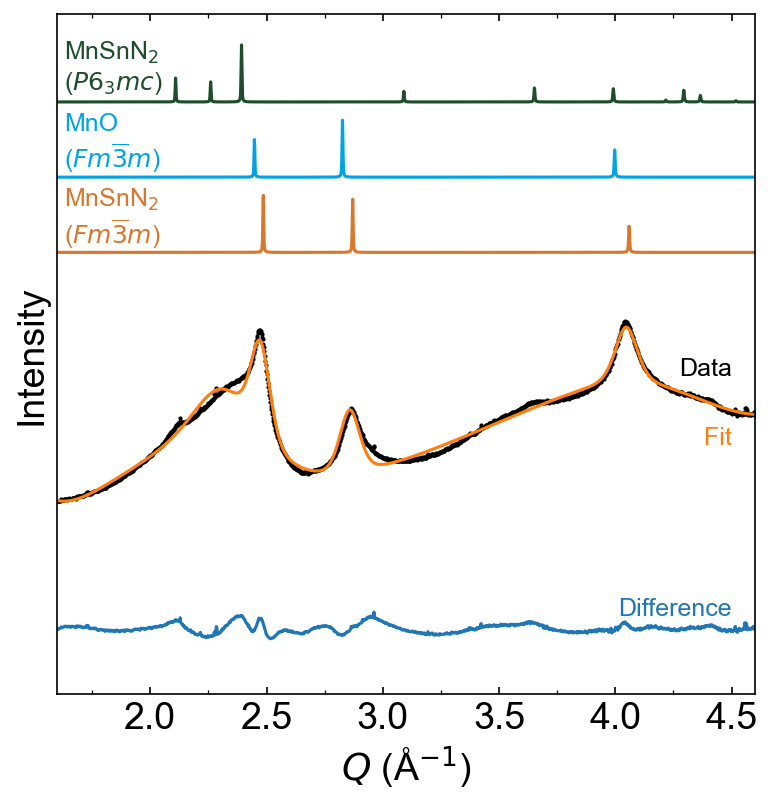}\\
  \caption{GIWAXS pattern for a sample at 68\% Mn/(Mn+Sn) and $T_\mathrm{dep}$ = 100~\textcelsius{} showing a rocksalt ($Fm\overline{3}m$) crystal structure ($a = 4.378$~\AA{}).}
  \label{fgr:rocksalt}
\end{figure}
A major finding in the thin film synthesis of \ce{MgSnN2} by Greenaway et al. was a wurtzite-to-rocksalt structural transition, and we find a small amount of evidence that a similar phenomenon occurs in the \ce{MnSnN2} system (Figure \ref{fgr:rocksalt}). In our studies of \ce{MnSnN2}, a single synchrotron GIWAXS  pattern revealed a rocksalt type structure at Mn/(Mn+Sn) = 68\%. The rocksalt appears under similar conditions needed for rocksalt \ce{MgSnN2} (Sn poor, low $T_\mathrm{dep}$). However, the refined lattice parameter for our \ce{MnSnN2} rock salt ($a = 4.378$~\AA{}) is very similar to the rocksalt lattice parameter for MnO ($a = 4.446$~\AA{}); the ternary is 1.5\% smaller than the binary. In contrast, rocksalt \ce{MgSnN2} ($a = 4.483$~\AA{}) is 6.3\% larger than MgO ($a = 4.217$~\AA{}). The presence of a rocksalt polymorph in the Mn-Sn-N system may merit further study.

\begin{figure}
    \centering
    \includegraphics[width = 0.5\textwidth]{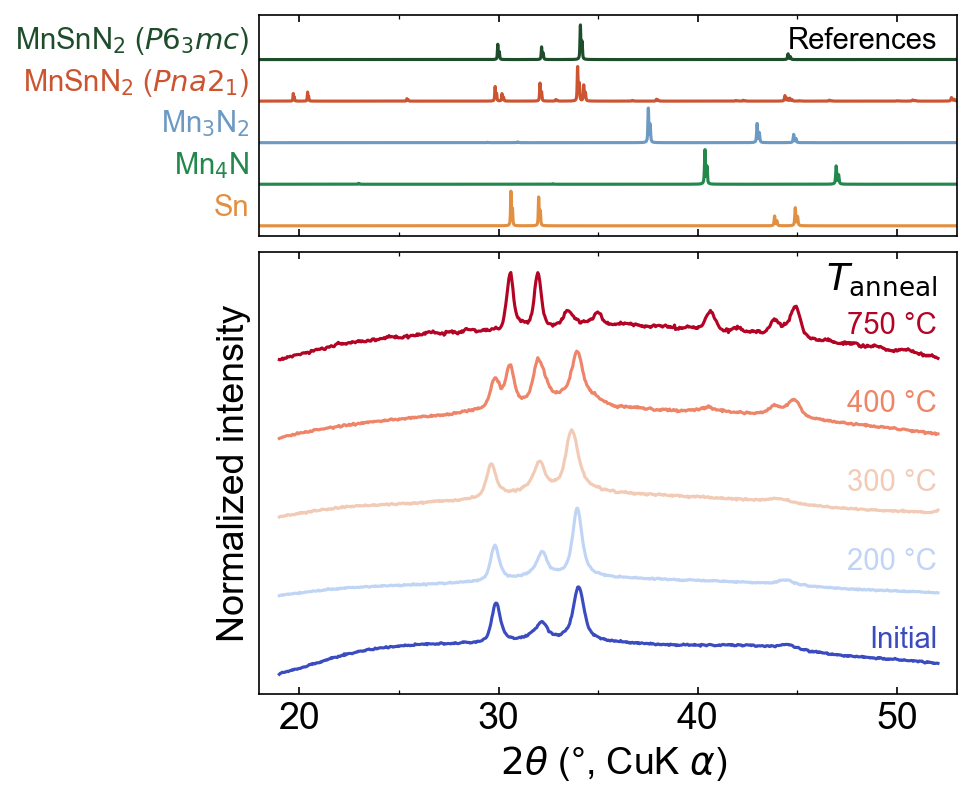}
    \caption{Rapid thermal annealing experiments show that a library row synthesized at $T_{\text{dep}} = 225$~\textcelsius{} is stable up to an annealing temperature of 300~\textcelsius{}.  Annealing experiments were conducted on individual rows from a library, first heated to 100~\textcelsius{} to drive off adsorbed water, then heated to the $T_{anneal}$ temperature for a 30~min anneal. Heating ramps for each step were conducted in 1 minute, and cooling occurred by turning the power to the furnace off.  The sample was enclosed in flowing \ce{N2} for the duration of the experiment. }
    \label{fig:RTA}
\end{figure}

Rapid thermal annealing experiments revealed that the \ce{MnSnN2} phase is stable against decomposition up to at least $T_{\mathrm{anneal}} =300$~\textcelsius{} for 30 minutes (Figure \ref{fig:RTA}).  At $T_{\text{anneal}} = 400$~\textcelsius{}, the ternary phase begins decomposing to Sn, \ce{Mn4N}, and \ce{N2}. This stability up to 300~\textcelsius{} suggests either i) that cation-disordered \ce{MnSnN2} is more thermodynamically stable that predicted for the cation-ordered phase (+0.038 eV/atom above the hull) or ii) that decomposition up to 300~\textcelsius{} is kinetically slow. 

\begin{figure}
    \centering
    \includegraphics[width = 0.5\textwidth]{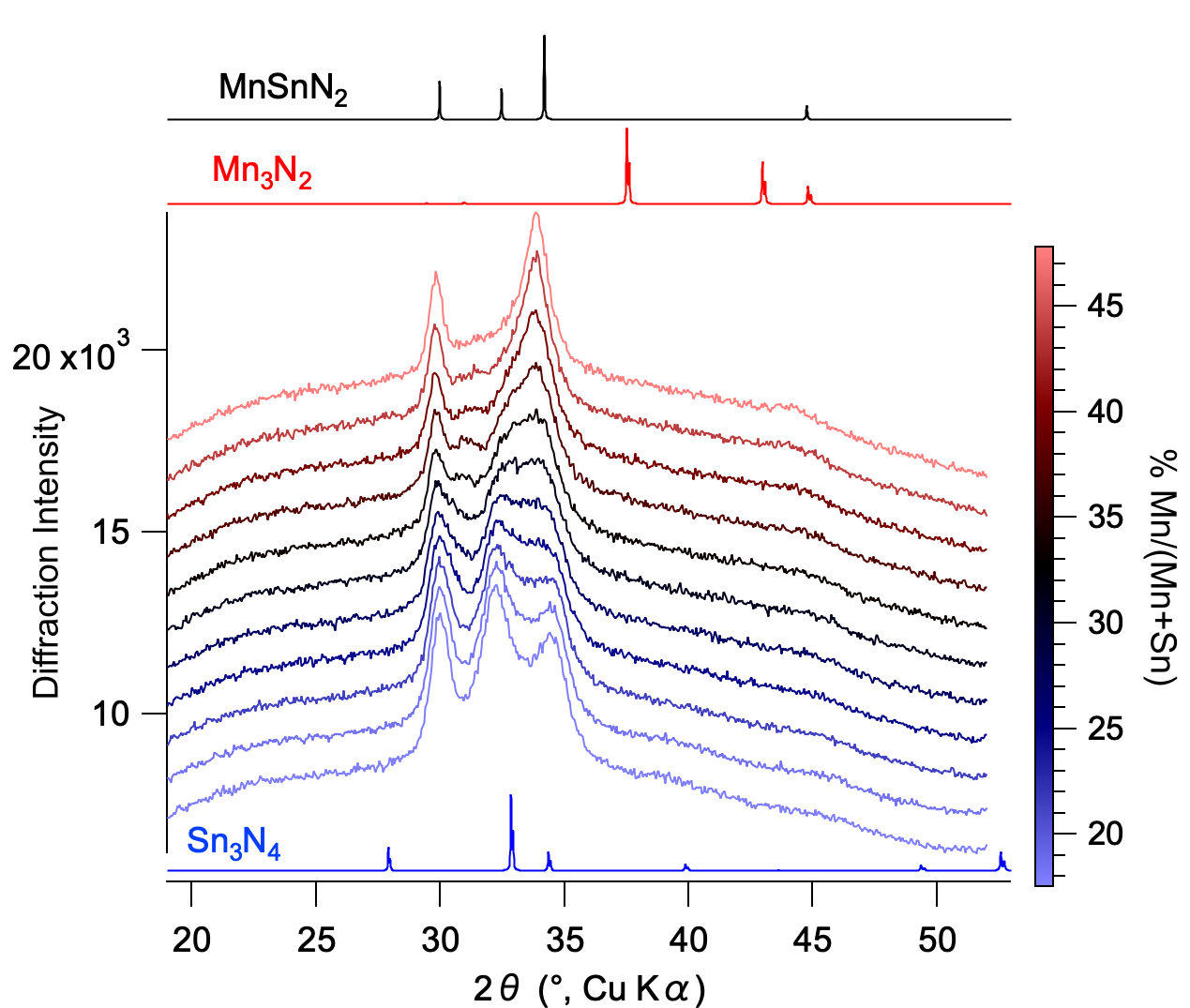}
    \caption{A library deposited under Mn-poor conditions at $T_{\text{dep}} = 225$~\textcelsius{} show the continued presence of wurtzite peaks. A structural change to the spinel \ce{Sn3N4} is not observed, although prior sputtering syntheses of this binary phase revealed poor crystallinity of binary Sn-N nitrides.\cite{caskey2016synthesisSn3N4}}
    \label{fig:Mn_poor_XRD}
\end{figure}

To assess the limits of off stoichiometry in the hexagonal \ce{MnSnN2} phase, one library was deposited with a lower power on the Mn target (54~W as opposed to the typical 108~W), leading to a Mn-poor film (Figure \ref{fig:Mn_poor_XRD}).  This Mn-poor film still exhibited wurtzite-like reflections down to approximately 18\% Mn/(Mn+Sn), limited by the edge of the library rather than by a structural change. This composition nominally corresponds to \ce{Mn_{0.36}Sn_{1.64}N2} (assuming a fixed anion content). For compositions with Mn/(Mn+Sn)$<50$\%, charge balancing must occur through either \ce{Sn^{2+}} substitution for \ce{Mn^{2+}} or metal vacancies. Although we did not explore this Mn-poor region further, this finding suggests that Mn-poor \ce{MnSnN2} could be of interest for dilute magnetic semiconductor studies. 

\clearpage
\section{Compositional characterization}
\subsection{RBS}
\begin{figure*}[h]
    \centering
    \includegraphics[width = 0.7\textwidth]{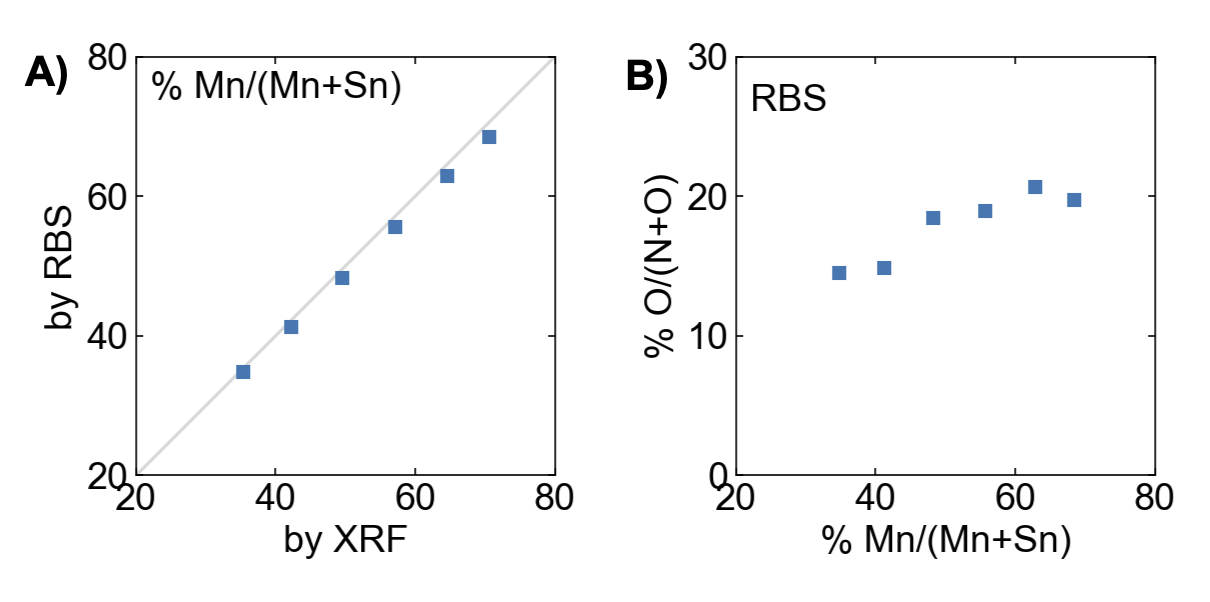}
    \caption{A) XRF measurements of \% Mn/(Mn+Sn) agree well with measurements by RBS. The grey trace represents perfect agreement between the techniques. Blue squares are the data. B) RBS measurements of \% O/(N+O) show that the proportion of oxygen content increases slightly with increasing Mn content.
    }
    \label{fig:RBS_merged}
\end{figure*}

\begin{figure*}[h]
    \centering
    \includegraphics[width = 0.7\textwidth]{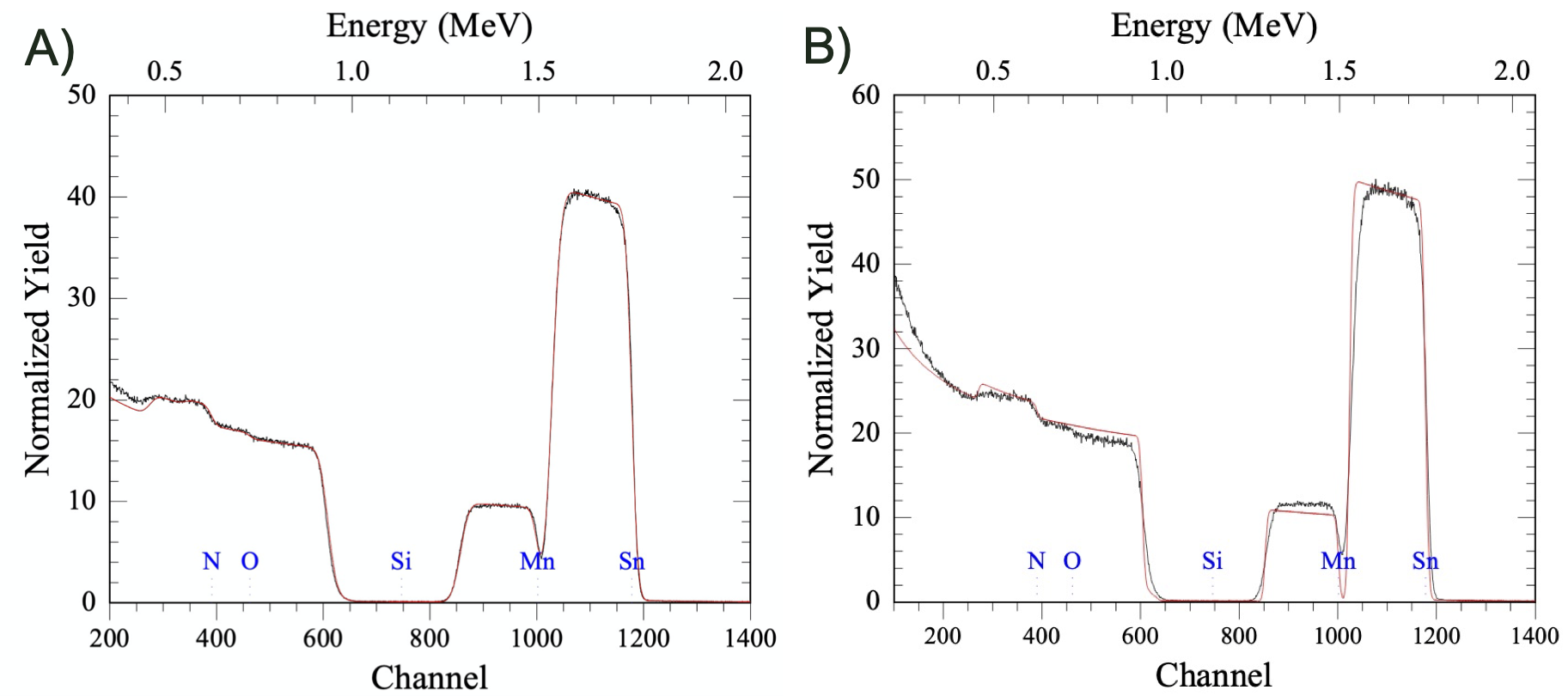}
    \caption{Representative RBS data (black traces) of one sample and fits (red traces) following different fitting procedures. A) Fit by freely varying a fuzz parameter, film thickness, and relative amounts of Mn, N, O, normalized to Sn = 1.  The minimization overfits the anion content, leading to an unreasonable ratio of (N+O)/(Mn+Sn) = 1.6 (expecting 1 for the wurtzite structure). B) Fitting the same data by refining thickness and Mn content and constraining N content to N = Mn+Sn (i.e., Sn = 1, Mn = 0.8, N = 1.8).  While the lack of a fuzz parameter leads to poor fitting of curvature, the constrained anion content still produces a reasonable fit to the intensity in the 0.5 MeV region.  Quantitative assessment of low-atomic number anion content is therefore tricky given the low signal and high background. However, these data show a non-zero oxygen content with a substantially larger proportion of nitrogen content. 
}
    \label{fig:RBS_fit_comparison}
\end{figure*}

RBS measurements for select samples corroborate the high-throughput XRF measurements done for each combinatorial film (Figure \ref{fig:RBS_merged}A).  Furthermore, RBS measurements identify the presence of oxygen in the \ce{MnSnN2} films (Figure \ref{fig:RBS_merged}B).  The convolution of the N and O signals in RBS with the signal from the substrate inhibit the quantitative relation of anion content (N+O) with cation content (Mn+Sn) (Figure \ref{fig:RBS_fit_comparison}), but the qualitative comparison of the trends in O/(N+O) as a function of cation content are still valid.  Oxygen content increases with increasing Mn content, consistent with the lower electronegativity of Mn (1.55 on the Pauling scale) compared to Sn (1.96). The width of the oxygen region of the RBS spectra is similar to the width of the other elemental regions, suggesting that oxygen is distributed throughout the film thickness rather than being isolated to the surface.
Oxygen substitution might help charge balance excess Mn via \ce{Mn_{1+$y$}Sn_{1-$y$}N_{2-$y$}O_$y$}, where $y$ is both the number of oxides substituting for nitrides and the number of \ce{Mn^{3+}} ions substituting for \ce{Sn^{4+}}. Alternatively, oxygen substitution could occur with exclusively \ce{Mn^{2+}}, following a formula of (MnO)$_x$(\ce{MnSnN2})$_{(1-x)}$, similar to the oxygen balancing seen in \ce{MgZrN2} thin films.\cite{bauers2019compositionMgZrN2}

\clearpage
\subsection{XANES}

In an attempt to identify the presence of \ce{Mn^{3+}} and assess \ce{Mn^{2+}}:\ce{Mn^{3+}} ratios, high-resolution Mn $L_3$-edge X-ray absorption near-edge spectroscopy (XANES) measurements were conducted. Measurements were carried out on thin films at room temperature at the Stanford Synchrotron Radiation Lightsource (SSRL), SLAC National Accelerator Laboratory, beamline 10-1. Samples were measured under high vacuum conditions of $\sim$2.7$\times10^{-6}$ Pa ($\sim$2$\times10^{-8}$ Torr) with a ring current of 500 mA. The synchrotron radiation was monochromatized using the beamline's 600~line/mm monochromator with entrance and exit slits of 35~$\mu$m. A transition edge sensor (TES) spectrometer \cite{TES} was used to collect resonant inelastic x-ray scattering (RIXS) planes with a resolution of 2~eV. The energy measured by the TES was calibrated by periodically measuring a reference sample of graphite, \ce{BN}, \ce{Fe2O3}, \ce{NiO}, and \ce{CuO}, which provide a stable set of emission lines. From the RIXS planes, we extracted the $L_{\alpha}$ and $L_{\beta}$ lines, leading to partial fluorescence yield XAS.

\begin{figure*}
    \centering
    \includegraphics[width=0.9\textwidth]{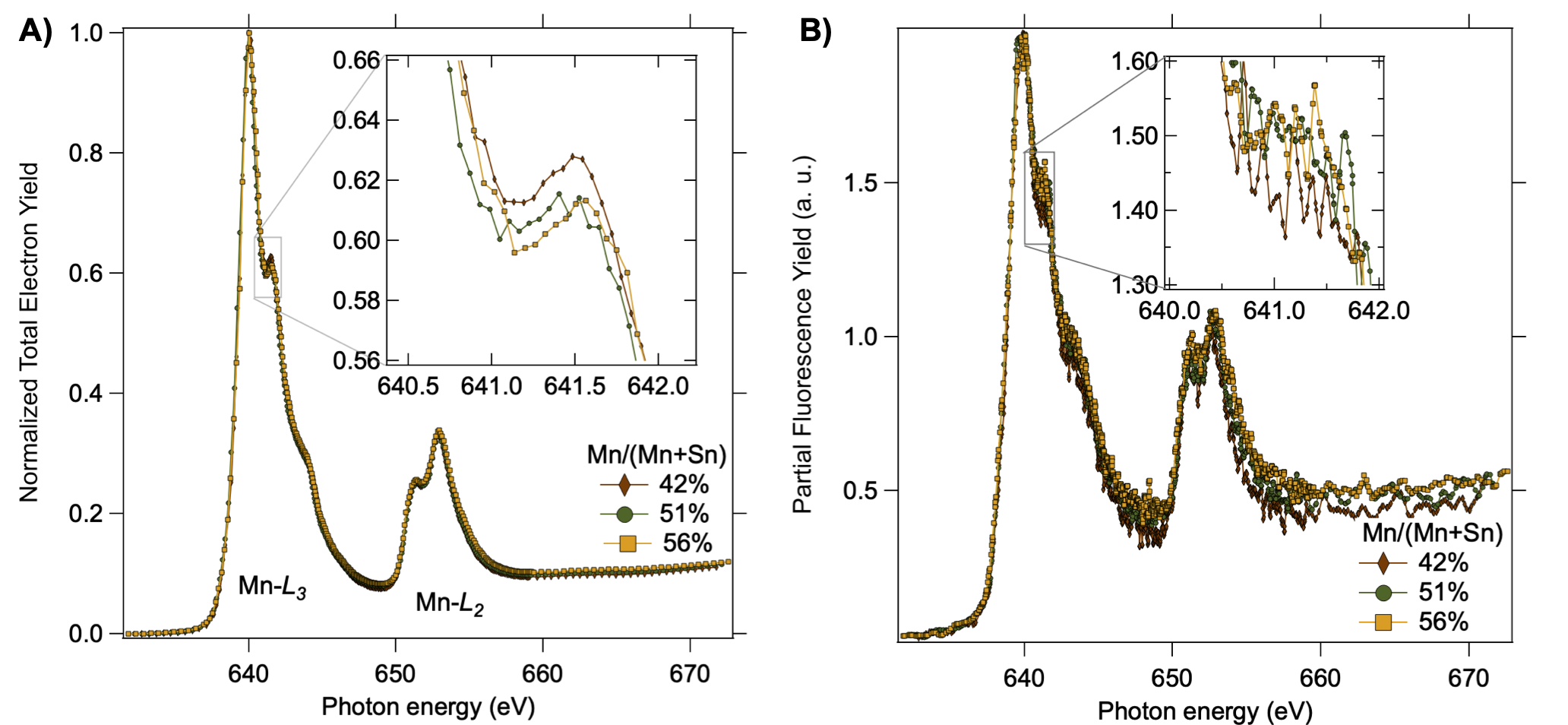}
    \caption{
    The XANES total electron yield (A) and partial fluorescence yield (B) of the Mn-\textit{L} edge shows that all three samples used for the magnetism measurements have similar spectral profiles. This profile matches \ce{Mn^{2+}} compounds in literature.\cite{niewa2002xas, cramer2002Mn_L_edge_XAS} Subtle differences can be seen on the peak shoulders of the L3 edge, but these shifts are near the noise level of the measurement. }
    \label{fig:xanes}
\end{figure*}

XANES measurements (Figure \ref{fig:xanes}) show consistent spectra across the measured composition range, so we are not able to identify a change in oxidation state.
There are subtle differences between the spectra, as highlighted in the insets, but these differences are hard to distinguish from noise. We observe no consistent trend with composition and no distinct features indicative of \ce{Mn^{3+}}. Instead, each the spectral profile is consistent with \ce{Mn^{2+}} spectra seen in literature.\cite{niewa2002xas, cramer2002Mn_L_edge_XAS}

\clearpage

\section{Magnetic properties}

\subsection{Background subtraction and data analysis}
\begin{figure}[h]
    \centering
    \includegraphics[width=0.35\textwidth]{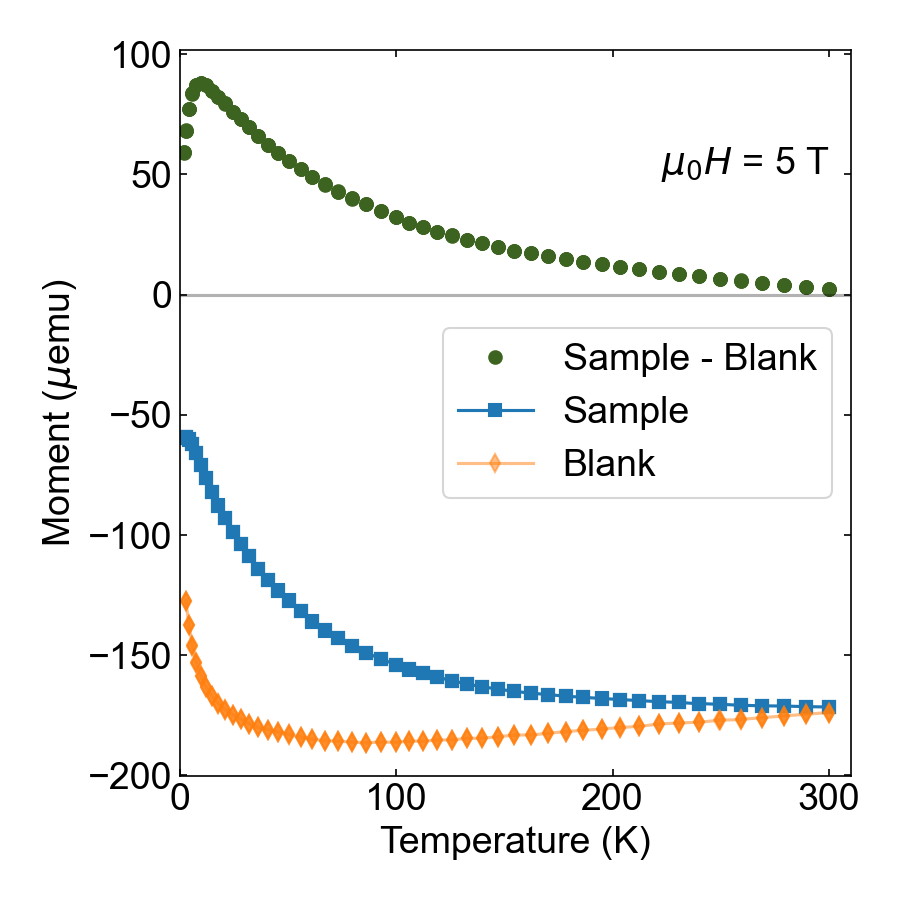}
    \caption{Background subtraction example for zero-field cooled temperature sweeps measured at $\mu_0H = 5$ T. The raw moments of the sample 
    and a blank substrate 
    are shown, along with the background-subtracted data showing magnetism from the film alone (sample $-$ blank).
    }
    \label{fig:backgroundSubtraction_MT}
\end{figure}

\begin{figure}[h]
    \centering
    \includegraphics[width=0.35\textwidth]{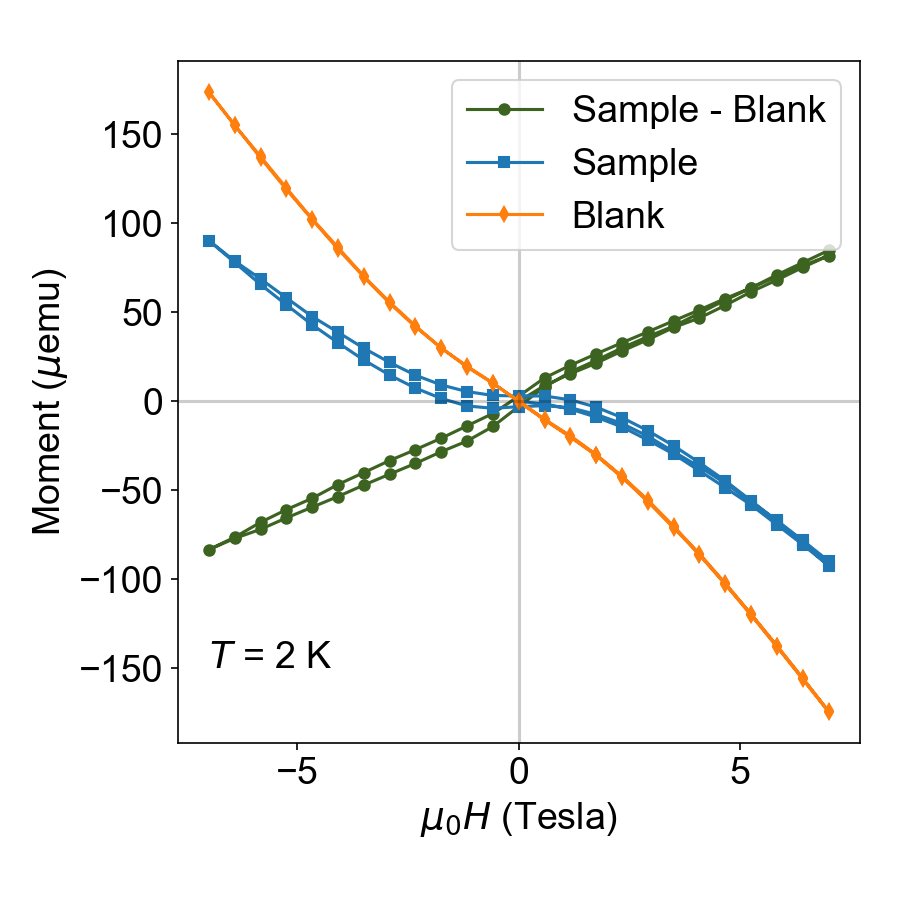}
    \caption{Background subtraction example for applied field sweeps measured at $T=2$ K. The raw moments of the sample 
    and a blank substrate 
    are shown, along with the background-subtracted data showing magnetism from the film alone (sample - blank)
    }
    \label{fig:backgroundSubtraction_MH}
\end{figure}

As noted in the main text, for each  magnetic measurement performed, an analogous blank substrate was measured with identical scan parameters, allowing for direct, point by point background subtraction.  These blanks were 5~mm $\times$ 5~mm squares of the substrate alone (100 nm \ce{SiO2} on pSi), taken from the same wafer batch used for the deposition.  
Figure \ref{fig:backgroundSubtraction_MT}--\ref{fig:backgroundSubtraction_MH} show examples of the background subtraction for temperature- and applied field-dependent measurements, respectively.  

Following background subtraction, the susceptibility ($\chi$) was calculated 
using the background-subtracted raw moment value described above, along with a film mass of 0.00004 g and the molar mass calculated as \ce{Mn_{2$x$} Sn_{2-2$x$}N2} (Table \ref{tab:df_sq_table}) where $x$ is the Mn fraction of cation content (Mn/[Mn+Sn]). The \% Mn/(Mn+Sn) values were determined by XRF on the individual squares. We note that the small mass of \ce{Mn_{2x}Sn_{2-2x}N2} present during the measurements and the several possible sources of error in the film mass calculation and background subtraction make direct, quantitative comparisons of the magnitude of $\chi$ between samples difficult. The value of 0.00004~g for a film mass is a reasonable approximation given the film surface area (25~mm$^2$), thickness, and composition.

\begin{table}[h!]
    \centering
\begin{tabular}{cccccc}
\toprule
\% Mn/(Mn+Sn) & Molar Mass (g/mol) & Thickness ($\mu$m) & $a$ (Å) & $c$ (Å) & $T^\mathrm{*}$ (K) \\\midrule
          42 &                212 &           0.305 & 3.449 & 5.667 &      6 \\
          51 &                201 &           0.304 & 3.441 & 5.562 &     10 \\
          56 &                194 &           0.302 & 3.438 & 5.536 &     12 \\
\bottomrule
\end{tabular}
    \caption{Summary of composition, thickness, hexagonal lattice parameters, and magnetic transition temperatures ($T^\mathrm{*}$) for the samples selected for magnetic measurements.}
    \label{tab:df_sq_table}
\end{table}

\clearpage
\subsection{Additional data}

\begin{figure*}
    \centering
    \includegraphics[width=0.9\textwidth]{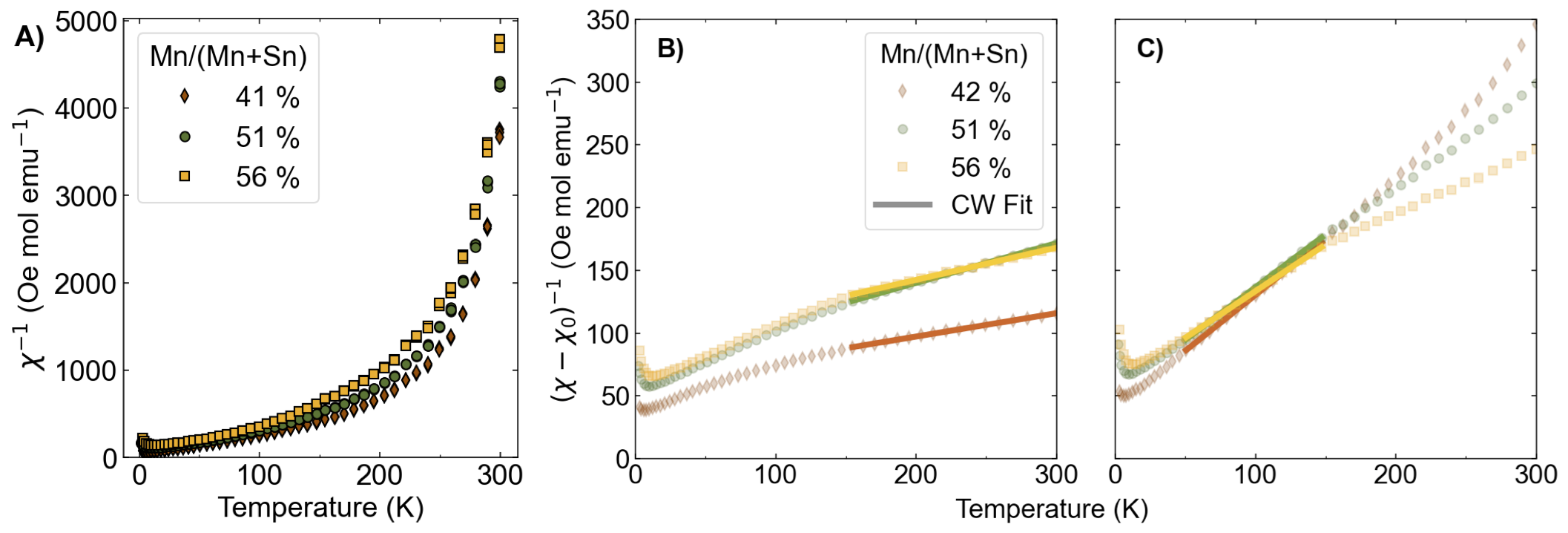}
    \caption{A) Inverse susceptibility plotted without a diamagnetic correction factor shows significant non-linearity. Inverse susceptibility data and Curie-Weiss fits performed with a diamagnetic correction ($\chi_0$) between B) 150--300 K C) 50--150K. The fits are shown as solid bars. Parameters $\Theta_{CW}$ and $\chi_0$ are shown in Table \ref{tab:CW}.
    C) 
    }
    \label{fig:CW_fits}
\end{figure*}

Inverse susceptibility data as a function of temperature show an approximately linear region above the $T^\mathrm{*}\sim$10~K transition temperatures for each composition, provided a diamagnetic correction ($\chi_0$) is applied (Figure \ref{fig:CW_fits}). Inverse $\chi$ without a diamagnetic correction exhibits significant non-linearity (Figure \ref{fig:CW_fits}A). Therefore, Curie-Weiss fits were performed on $(\chi - \chi_0)^{-1}$ from 150--300 K, as shown in Figure \ref{fig:CW_fits}B. As the inverse $\chi$ plots exhibited a second linear region (50--150~K), we also applied Curie-Weiss analysis to this lower temperature region (Figure \ref{fig:CW_fits}C), which resulted in a smaller magnitude $\Theta_{CW}$ and a smaller magnitude $\chi_0$ than the high temperature fits (Table \ref{tab:CW}). Weiss temperatures from both fit ranges indicate antiferromagnetic (AFM) correlations. The change in slope may indicate a change in the energy scale of the dominant interactions in different temperature regions, or the emergence of competing interactions, although further investigation will be required to elucidate these subtle differences.
We note that the uncertainty in background subtraction and film mass discussed above and in the main text inhibits quantitative extraction of Curie constants ($C$) or effective moments ($\mu_\text{eff}$) from the Curie-Weiss fits. 

\begin{table}[h!]
\centering
    \caption{Weiss temperatures ($\Theta_{CW}$) extracted from Curie-Weiss fits performed from 50--150K and 150--300 K with the diamagnetic corrections ($\chi_0$) listed.}
\begin{tabular}{c|cc|cc} 
\toprule
\multirow{2}{*}{\% Mn/(Mn+Sn) } & \multicolumn{2}{c|}{50--150 K}  & \multicolumn{2}{c}{150--300 K}\\ 
&  $\Theta_{CW}$ (K) &  $\chi_0$ (emu/Oe mol) & $\Theta_{CW}$ (K) &  $\chi_0$ (emu/Oe mol)\\ 
\hline
          42 &         -47 &  -0.0025 &         -324 &   -0.0083 \\
          51 &         -63 &  -0.0030 &         -243 &   -0.0055 \\
          56 &         -79 &  -0.0038 &         -356 &   -0.0057 \\
\bottomrule
\end{tabular}
    \label{tab:CW}
\end{table}

\begin{figure}[ht!]
    \centering
\includegraphics[width=0.35\textwidth]{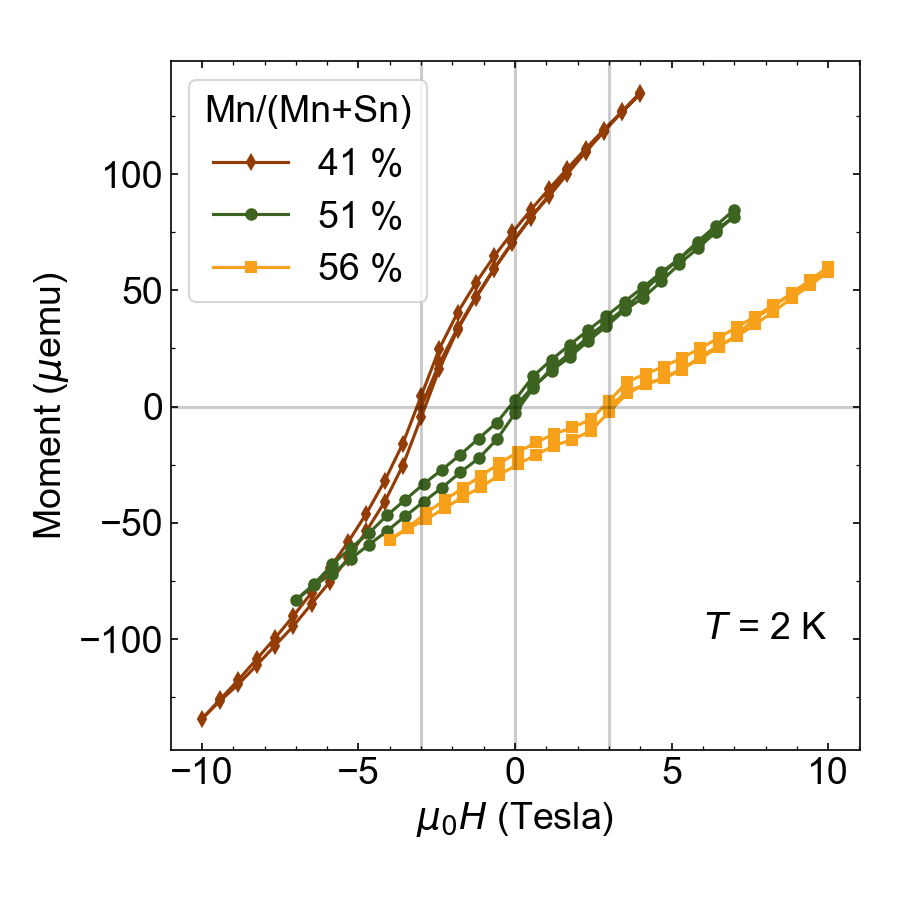}\\
  \caption{Isothermal magnetic moment measurements at 2~K of \ce{MnSnN2} samples show decreasing moment with increasing Mn content. Traces are offset horizontally by $\mu_0H = 3$ T for clarity (grey lines guide the eye to each origin). The substrate contribution has been subtracted. } 
  \label{fgr:magnetism_MH}
\end{figure}

Figure \ref{fgr:magnetism_MH} shows the background-subtracted moment as a function of applied field measured at $T=2$ K for all samples. This low-temperature behavior (below the transition temperature) is consistent with an AFM ground state across the composition series. The small hysteresis and net ferromagnetic moment noted in the Mn/(Mn+Sn) = 51\% sample (Figure \ref{fig:Magnetism_MH_2K_40K} in the main text) are also present across the measured composition range, supporting our assignment of this moment arising from canting or defects. The samples do not exhibit a plateau in moment, suggesting that the magnetism is not saturated at the limit of the instrument ($\mu_0H = 7$ T). The moment at $\mu_0H = 7$ T decreases with increasing Mn/(Mn+Sn) content, attributable to either the substitution of \ce{Mn^{3+}} for \ce{Mn^{2+}} or uncertainty in the mass used for normalization. Also, the low-field slope changes with increasing Mn/(Mn+Sn): at Mn/(Mn+Sn) = 56\%, it appears to have two components, while at Mn/(Mn+Sn) = 42\% there is only one component. This behavior may be related to uncompensated moments from \ce{Mn^{3+}}.

\clearpage
\section{Optoelectronic measurements}
\begin{figure*}[hb]
\centering
\includegraphics[width=\textwidth]{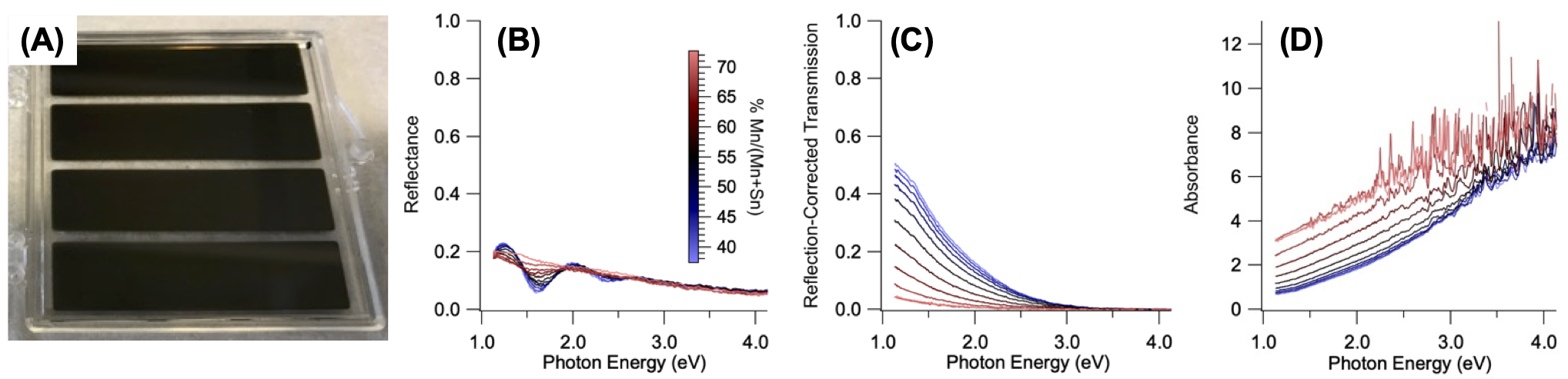}\\
  \caption{A) Photo of a library deposited on a 2x2" Eagle XG glass substrate. The library was deposited at a set point of 250~\textcelsius{}, which corresponds to a substrate temperature of $<225$~\textcelsius{} as evidenced by the lack of decomposition of the \ce{MnSnN2} phase (confirmed by XRD). The film is dark brown in color.  B) Reflectance \textit{R} and C) Reflection-corrected transmittance \textit{T} were used to calculate the absorbance via Absorbance $A= \ln{\frac{T}{1-R}}$, as implemented in Combigor.\cite{talley2019combigor}. D) The absorbance is high across measurable range. The colorbar shown in panel B is used for panels C and D as well.}
  \label{fgr:uvvis}
\end{figure*}

In addition to the ellipsometry measurements and modeling reported in the main text (Figure \ref{fgr:optoelectronics}C), we conducted transmission and reflection UV-vis measurements on select substrates deposited on EXG, a transparent substrate (Figure \ref{fgr:uvvis}). 
Transmission ($T$) and reflectance ($R$) spectra were collected in the UV-Vis-NIR spectral ranges (300–1100~nm) using a home-built thin film optical spectroscopy system equipped with deuterium and tungsten/halogen light sources and a Si detector array. The collected spectra were then used to calculate absorbance using the relation Absorbance, $A = \ln{[T/(1 - R)]}$.
Unfortunately, the UV-vis spectrometer did not reach a sufficiently low energy to observe the absorption onset (1~eV, 1200~nm). However, these data support a low energy absorption onset without the need for modeling of the spectra (as needed for ellipsometry). 

\begin{figure}[h]
    \centering
\includegraphics[width=0.35\textwidth]{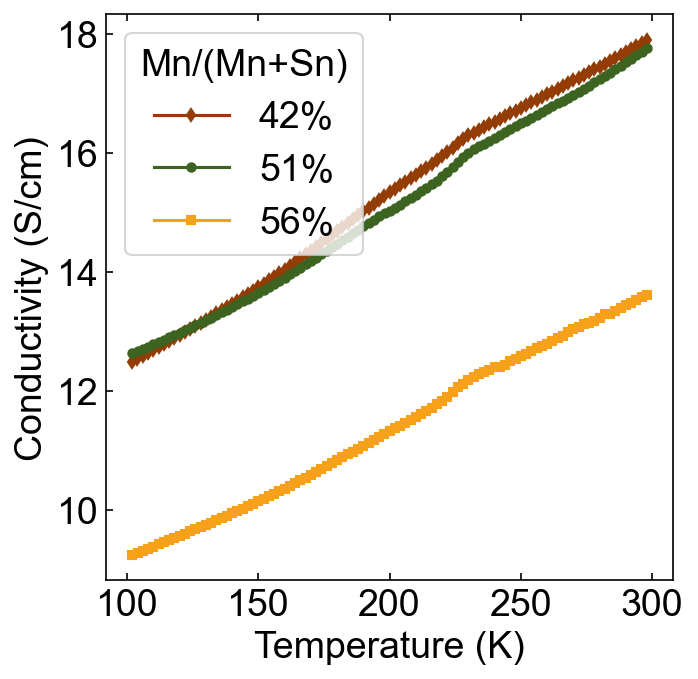}\\
  \caption{Conductivity measurements on 5~mm $\times$ 5~mm squares scribed from a library deposited at $T_{\text{dep}} = 225$~\textcelsius{} on Si with 100 nm \ce{SiO2}. }
  \label{fgr:Conductivity_v_temperature_raw}
\end{figure}

In addition to measuring electrical conductivity of a film on EXG (Figure \ref{fgr:optoelectronics}), we also conducted temperature dependent electrical measurements on select samples deposited on an insulating 100 nm of \ce{SiO2} on Si (Figure \ref{fgr:Conductivity_v_temperature_raw}). These measurements were consistent with our observations for the sample on EXG (Figure \ref{fgr:optoelectronics}B), showing increasing conductivity with increasing temperature. However, these blanket films on 100 nm of \ce{SiO2} have a tendency to short across such a narrow insulating layer if cleaved (as for these samples). To ensure we were exclusively measuring conductivity of the film, rather than the Si substrate, we repeated the measurement on EXG and report that in the main body of the text. 

\end{document}